# Electromagnetic Composites: from Effective Medium Theories to Metamaterials

Faxiang Qin[1,a),b)], Mengyue Peng[1,a)], Diana Estevez[1], Christian Brosseau[2]

[1]*Institute for Composite Science Innovation (InCSI), School of Materials Science and Engineering, Zhejiang University, Hangzhou, 310027, China*

[2]*Laboratoire d'Electronique et Systèmes de Télécommunications, Université de Bretagne Occidentale, 6 avenue Le Gorgeu, B. P. 809, 29285 Brest Cedex, France*

---

a) Equally contributing authors
b) Corresponding author: faxiangqin@zju.edu.cn

# ABSTRACT

Electromagnetic (EM) composites have stimulated tremendous fundamental and practical interests owing to their flexible electromagnetic properties and extensive potential engineering applications. Hence, it is necessary to systematically understand the physical mechanisms and design principles controlling EM composites. In this tutorial, we first provide an overview of the basic theory of electromagnetism about electromagnetic constitutive parameters that can represent the electromagnetic properties of materials. We show how this corpus allows a consistent construction of effective medium theories and allows for numerical simulation of EM composites to deal with structure-property relationships. We then discuss the influence of spatial dispersion of shaped inclusions in the material medium on the EM properties of composites, which has not been systematically illustrated in the context of this interdisciplinary topic. Next, artificial composites or metamaterials with peculiar properties not readily available in nature are highlighted with particular emphasis on the control of the EM interaction with composites. We conclude by discussing appropriate methods of electromagnetic measurement and practical aspects for implementing composites for specific applications are described. Overall, this tutorial will serve the purpose of introducing the basics and applications of electromagnetic composites to newcomers in this field. It is also anticipated that researchers from different backgrounds including materials science, optics, and electrical engineering can communicate to each other with the same language when dealing with this interdisciplinary subject and further push forward this advancement from fundamental science to technological applications.

# I. INTRODUCTION

Since the discovery of the connection between electric field and magnetic field by Ampere and Faraday, which was later formulated into the famous Maxwell equations by James Clerk Maxwell in the 19$^{th}$ century, the exploration of electromagnetics has never been stopped in both science and technology fronts. In essence, Maxwell equations lay the foundation for human beings understanding and creating the contemporary electromagnetic world as it sees only charges and electrical currents but not at the materials level. Today, there are numerous textbooks discussing electromagnetics and electrodynamics, but tutorials on electromagnetic materials are few. However, as the significance of electromagnetics and relevant applications lie in a complete understanding of how electromagnetic wave interacts with the matter, it is essential to have a full knowledge of the electromagnetic materials from a materials science perspective, i.e., how charges, currents and their corresponding responses to the electromagnetic wave constitute a material property that can be modulated. Thus, the main purpose of this tutorial is to elucidate the relationships between the electromagnetic properties of much scientific and technical interest and the structural parameters across multiple scales. Only in this way can one approach the optimized design of the electromagnetic materials that can meet specific applicational requirements. Also, we place our major efforts on the electromagnetic composites as they are arguably one of the most important electromagnetic materials since their collective responses of such heterogeneous materials offer a multitude of possibilities to tailor the properties in a wide range and very well controlled fashion. They are also likely to invoke versatile wave-matter interaction physics. Most significantly, electromagnetic composites form a type of composite materials constituted by electromagnetic inclusions dispersed or arranged in a continuous matrix following a prescribed manner to yield meaningful electromagnetic responses. This last statement raises important questions about the role of inclusions, matrix, their interface, and the dispersion or arrangement pattern for inclusions in the matrix which eventually determine the final electromagnetic properties of composites. Indeed, composites are one of the most intriguing materials that can afford the functionality following a bottom-up paradigm, instead of the top-down mode, as the fillers of composites can serve as an augmented atom (in analogy to a real atom's avatar, i.e., a scaled-up atom or a meta-atom in a metaverse) at researcher's disposal in a convenient scale for manipulating the wave-matter interactions to yield architectures in favor of prescribed properties. As such, the electromagnetic composites align with the 'bottom-up' trend to give a high efficiency of the material and structure to take on precise functionality as against to the top-down approach to receive functionality only at the system level without much control on the lower hierarchy components which may be redundant or at least not optimized. In light of all the importance of electromagnetic composites as described above, we write this tutorial to introduce to the graduate students the condensed basics and applications of electromagnetic composites with an emphasis on the materials science perspective. The rest of the tutorial is organized as follows. We start with Sec. II by going through the electromagnetic constitutive parameters used to characterize the electromagnetic properties of materials and their frequency dispersion characteristics. Those well-informed readers who are familiar with electromagnetics are kindly advised to skip this section which is set up for the students who are new to this field. We then move to Sec. III to show how to predict the composite property with the knowledge of composite microstructures, i.e., establish the structure-property relationship. For this purpose, special emphasis is given to the effective medium theories and numerical models of electromagnetic composites. Subsequently, we discuss more generally the role of the spatial pattern of inclusions on the electromagnetic response in Sec. IV. Next, we shift our attention from passive understanding of the wave-matter interaction phenomenon as the filler is way smaller than the wavelength to active design with the filler organized at the subwavelength scale. A new paradigm to investigate electromagnetic composites is then revealed in Sec. V, dealing with a special class of innovative EM composites materials, metamaterials, with peculiar EM properties. Sec. VI discussed the advantages and disadvantages of electromagnetic measurement approaches, and technological applications of electromagnetic composites such as antennae, cloaking, and structural health monitoring. The whole tutorial is concluded in the last section along with a brief outlook.

## II. ELECTROMAGNETIC CONSTITUTIVE PARAMETERS

### A. Definition and fundamental sources of electromagnetic constitutive parameters

The starting point for analyzing the macroscopic response of a material to its interaction with an electromagnetic wave field relies on its constitutive parameters: permittivity $\varepsilon$ and magnetic permeability $\mu$, referring to the polarization and magnetization responses, respectively. $\varepsilon$ and $\mu$ for most materials are complex, and frequency-dependent parameters. Such dependence is referred to as dielectric and magnetic dispersions.

*1. Complex permittivity ε*

For a perfect dielectric material, $\varepsilon$ is characterized by a real number, but in most materials, $\varepsilon$ is described by a complex number with its imaginary part related to loss. Thus, $\varepsilon$ is called absolute complex permittivity and characterizes the response of the material to the varying electrical field

$$\varepsilon = \varepsilon' - j\varepsilon''. \tag{1}$$

The relative complex permittivity is defined by the ratio of $\varepsilon$ to the permittivity of free space $\varepsilon_o$ (8.854 x 10$^{-12}$ F/m)[1]

$$\varepsilon_r = \frac{\varepsilon}{\varepsilon_o} = \left(\frac{\varepsilon'}{\varepsilon_o}\right) - j\left(\frac{\varepsilon''}{\varepsilon_o}\right) = \varepsilon_r' - j\varepsilon_r'', \tag{2}$$

where $\varepsilon_r'$ is the real part of the relative complex permittivity characterizing the polarization property and stored energy within the material; $\varepsilon_r''$ is the imaginary part of the relative complex permittivity and polarization loss representing the dissipation/loss of energy within the material. When $\varepsilon_r$ is expressed in a vector form, the $\varepsilon_r'$ and $\varepsilon_r''$ components of the equivalent circuit must have a phase difference of 90° and the vector sum forms the angle $\delta$ with the real axis $\varepsilon_r'$ (Fig. 1a).[2] The tangent of this angle $\tan\delta_e$ (dielectric tangent loss) quantifies the dissipation of electrical energy due to processes such as relaxation, resonance, and loss from the delay in the response to the applied electric field.[3] Loss tangent can be calculated as

$$\tan\delta_e = \frac{\varepsilon_r^{''}}{\varepsilon_r^{'}}. \qquad (3)$$

Combining Eq. (2) and Eq. (3), the following relation is obtained

$$\varepsilon_r = \varepsilon_r^{'} - j\varepsilon_r^{''} = \varepsilon_r^{'}(1 - j\varepsilon_r^{''}/\varepsilon_r^{'}) = \varepsilon_r^{'}(1 - j\tan\delta_e). \qquad (4)$$

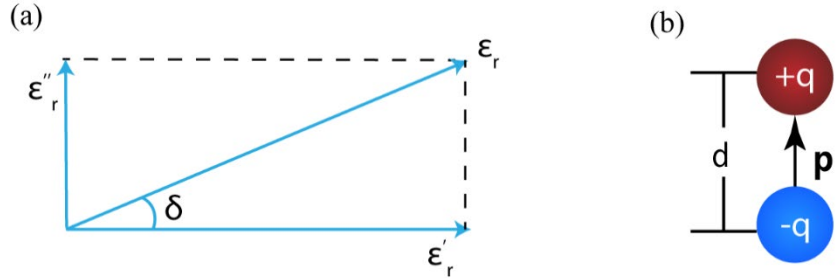

FIG. 1. (a) Vector diagram of the relative complex dielectric permittivity $\varepsilon_r$. (b) Schematic representation of a dipole formed by two charges, each of magnitude $q$, separated by distance $d$.

The fundamental origin of permittivity corresponds to the presence of electric charges, physical separation of positive and negative charges, and/or differential charge density.[4] Isolated charges can generate an electric field and similarly "dipolar/multipolar" fields are produced by assemblies of positive and negative charges placed at some distance. Permittivity thus relates to charge separation and such separation produces electric dipoles and multipoles. The dipole is characterized by an electric moment vector *p* pointing from the negative towards the positive charge with a magnitude equal to the product of charge and charge distance *p=qd* (Fig. 1b). Dipole moments generally form when the material is subjected to an external electric field, the charges move in response to the field and generate an additional internal electric field.[4] This induced field is represented by the polarization vector *P* which sums up all dipole contributions and takes a direction which is determined by the vector sum of individual dipole moments. The total internal field is thus the sum of the external electric field and the polarization field. Such sum corresponds to the electric displacement $\boldsymbol{D}$ which can be expressed as[5]

$$\boldsymbol{D} = \varepsilon_o \boldsymbol{E} + \boldsymbol{P}. \tag{5}$$

By considering a linear media, the polarization is proportional to the external electric field $\boldsymbol{E}$, electric susceptibility $\chi_e$, and free space permittivity $\varepsilon_o$, i.e. $\boldsymbol{P} = \varepsilon_o \chi_e \boldsymbol{E}$. Then from Eq. (5), a direct relationship between the displacement field and fields, we have[4]

$$\boldsymbol{D} = \varepsilon_o \boldsymbol{E} + \boldsymbol{P} = \varepsilon_o \boldsymbol{E} + \varepsilon_o \chi_e \boldsymbol{E} = \varepsilon_o \boldsymbol{E}(1 + \chi_e) = \varepsilon_o \varepsilon_r \boldsymbol{E}. \tag{6}$$

Thus, permittivity $\varepsilon$ is related to the electric susceptibility $\chi_e$, free space permittivity $\varepsilon_o$ and the dimensionless relative permittivity $\varepsilon_r$, giving $\varepsilon = \varepsilon_o(1 + \chi_e) = \varepsilon_o \varepsilon_r$.

### *2. Complex permeability μ*

The magnetic permeability $\mu$ can be related to the free space value $\mu_o = 4\pi \times 10^{-7}\,\text{H/m}$ in a similar fashion to that used for the permittivity

$$\mu = \mu_r \mu_o, \tag{7}$$

where $\mu_r$ denotes the relative permeability which is generally a complex number and can be expressed as

$$\mu_r = \frac{\mu}{\mu_o} = \left(\frac{\mu'}{\mu_o}\right) - j\left(\frac{\mu''}{\mu_o}\right) = \mu_r' - j\mu_r''. \tag{8}$$

where $\mu_r'$ is known as the real part of the relative permeability which represents the material storage capacity of the magnetic field, whereas the imaginary part $\mu_r''$ is related to losses and power dissipation due to the magnetic field. In addition, the real part can be related to inductance and the imaginary part to resistance.[6] Like permittivity, there is also a magnetic loss factor or power factor based on the angle between $\mu_r'$ and

$\mu_r^"$

$$\tan \delta_m = \frac{\mu_r^"}{\mu_r^'}, \quad (9)$$

where $\tan \delta_m$ represents the energy loss due to phase delay between the applied and induced magnetic fields. Magnetic losses originate from hysteresis losses (related to magnetic flux density), eddy current losses (related to frequency), and residual losses (associated with magnetic domain wall and spin rotational resonances).[7] From Eqs. (8) and (9), we have

$$\mu_r = \mu_r^' - j\mu_r^" = \mu_r^'(1 - j\mu_r^"/\mu_r^') = \mu_r^'(1 - j\tan\delta_m). \quad (10)$$

As was mentioned above, electric fields generate from isolated charges or charge assemblies, but in the case of the magnetic field, the source originates from currents. A macroscopic source exists if the material has finite electrical conductivity $\sigma$. If an electric field is applied between two points A and B, the free charges will move between these points and generate a free current $J_f$. This current produces a magnetic field whose magnitude is proportional to the current and which direction is defined by Lenz's law and right-hand rule (Fig. 2a). The free current is equal to $J_f = \sigma E$.[4]

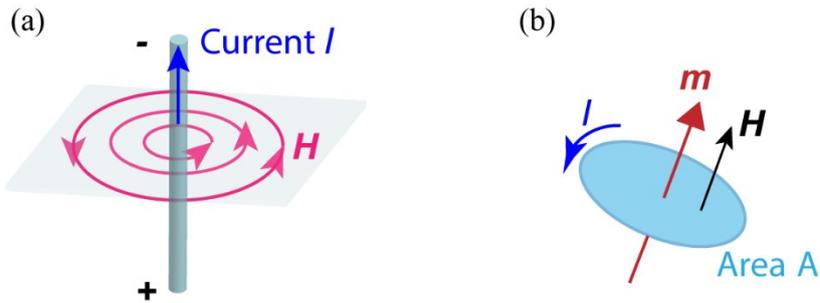

FIG. 2. Main sources of magnetic field (a) linear current; (b) magnetic dipole. The magnetic dipole $m$ is the product of the current $I$ and area $A$ with the direction shown in the figure.

The microscopic origin of the magnetic field is related to bound orbiting particles or spinning particles around a subatomic, atomic, or molecular axis. Magnetic materials, possess molecular and crystalline electronic configurations that produce a net alignment of spin and orbital momentum vectors.[4] Therefore, a tiny current loop behaves as a tiny magnet technically called a magnetic dipole (Fig. 2b). The product of the current passing through the material and the area circumscribed by the moving charge produces a magnetic dipole moment with a magnitude of $m = iA$, a direction perpendicular to the area and determined by the right-hand rule. The moment and associated magnetic field $H$ (magnitude is proportional to the current magnitude) are parallel to each other. Since the quantity of charge flowing past a point divided by time is current, and time is related to angular velocity, the magnetic moment is also related to the angular velocity. If a significant number of spins and/or orbital moments generate a vector alignment, the vector sum produces a net magnetization $M$. Similar to the case of electric polarization, the magnetization couples with the external applied field $B$ to generate an additional internal magnetic field $H$. Analogous to equations relating electric field, displacement and polarization, $B$, $H$, and $M$ are interrelated[4,5]

$$B = \mu_o(H + M) = \mu_o(H + \chi_m H) = \mu_o H(1 + \chi_m) = \mu_o \mu_r H, \qquad (11)$$

where $\chi_m$ corresponds to the magnetic susceptibility. Moreover, if there are internal geometrical variations in magnetization (e.g. if any electric dipoles changed their separation, orientation, or position as a function of time), an additional current is produced: $\nabla \times M$ (where $\nabla \times$ represents the curl operation). Thus, all current sources result in

$$\mu_o J_m = \mu_o \sigma E + \mu_o \frac{\partial P}{\partial t} + \mu_o \nabla \times M, \qquad (12)$$

where $\frac{\partial P}{\partial t}$ denotes the time variation in polarization.

**B. Dispersion relations of real and imaginary parts of permittivity (permeability)**
We furthermore observe that the real and imaginary parts of the permittivity and magnetic permeability of electromagnetic media cannot take independent values but are related by the Kramers-Kronig (KK) relations.[8-10] We also note that the KK relations connect the real and imaginary parts of causal response functions that are analytic in the upper half complex plane. The time dependence of polarization and magnetization follows the excitation field in time and is expressed as a function of the magnitude and frequency dispersion of permittivity-electric susceptibility $\varepsilon(\omega)-1=\chi_e(\omega)$ and/or permeability-magnetic susceptibility $\mu(\omega)-1=\chi_m(\omega)$. A typical form of KK relations is expressed in Eq. (13) where the integrals only involve positive frequencies; $\beta(\omega)=\beta'(\omega)+j\beta''(\omega)$ can be either dielectric or magnetic susceptibility[4]

$$\beta'(\omega)=\frac{2}{\pi}\mathcal{P}\int_0^\infty \frac{\eta\beta''(\eta)}{\eta^2-\omega^2}d\eta, \text{ and } \beta''(\omega)=-\frac{2\omega}{\pi}\mathcal{P}\int_0^\infty \frac{\beta'(\eta)-1}{\eta^2-\omega^2}d\eta. \qquad (13)$$

The $\mathcal{P}$ before the integral indicates the use of the Cauchy principle value of the integral, which is needed due to the singularity that the integral contains. Considering permittivity dispersion in the frequency region of polar relaxation with corresponding relaxation time $\tau$, the real and imaginary parts of permittivity-susceptibility read[4]

$$\varepsilon'(\omega)-1=\chi_e'(\omega)=\frac{\varepsilon(0)-\varepsilon(\infty)}{1+(\omega\tau)^2}=\frac{\chi_{e,DC}}{1+(\omega\tau)^2},$$

$$\varepsilon''(\omega)=\chi_e''(\omega)=\frac{(\varepsilon_{DC}-1)\omega\tau}{1+(\omega\tau)^2}=\frac{\chi_{e,DC}\omega\tau}{1+(\omega\tau)^2}. \qquad (14)$$

In this case, the real part can be determined from the imaginary part by direct substitution in the KK relations and evaluating the corresponding integral results in

$$\chi_e'(\omega)=\frac{2}{\pi}\mathcal{P}\int_0^\infty \frac{\nu\chi_{e,DC}\nu\tau}{(1+\nu^2\tau^2)(\nu^2-\omega^2)}d\nu=\frac{2\chi_{e,DC}}{\pi\tau^2}\mathcal{P}\int_0^\infty \frac{\nu^2\tau}{(\nu+j\tau)(\nu-j\tau)(\nu+\omega)(\nu-\omega)}d\nu. \qquad (15)$$

Here, $\mathcal{P}$ is determined by the sum of residues at the poles of the integral in the upper half complex plane, i.e. at $\nu=\omega$ and $j\tau^{-1}$. Therefore Eq. (15) reads

$$\chi_e'(\omega)=\frac{2\chi_{e,DC}}{\pi^2}\pi j\left\{\lim_{\nu\to\omega}(\nu-\omega)Integ(\nu,\omega,\tau)+\lim_{\nu\to j\tau^{-1}}(\nu-j\tau)Integ(\nu,\omega,\tau)\right\},$$

where $Integ(\nu,\omega,\tau)=\frac{\nu^2\tau}{(\nu+j\tau)(\nu-j\tau)(\nu+\omega)(\nu-\omega)}$. Evaluating the real part at the pole $\nu=j\tau^{-1}$ results in

$$\chi_e'(\omega)=\frac{2j}{\tau^2}\frac{\chi_{e,DC}(j\tau^{-1})^2\tau}{(2j\tau^{-1})(-\tau^{-2}-\omega^2)}=\frac{1}{\tau^4}\frac{-\chi_{e,DC}\tau^2}{(-\tau^{-2}-\omega^2)}=\frac{\chi_{e,DC}}{(1+\tau^2\omega^2)}. \qquad (16)$$

Therefore, the real part of permittivity $\varepsilon'$ in Eq. (14) is recovered by solving the integral.

Hence, in principle if either the real or imaginary part of permittivity/permeability is measured over the complete frequency range, the other unknown part can be calculated using the KK relations.[11] In addition, due to the causality constraint, models for frequency-dependent permittivity or permeability must satisfy the KK relations. For an effective theorist, getting the real (imaginary) part from its (imaginary (real)) part seems trivial. However, since the integral relationships have lower and upper frequency limits of 0 and ∞, the application of KK relations to measured data is in fact not so straightforward. Calculating either real part from imaginary part or vice versa requires very dense data of one quantity over a large bandwidth. With the development of the network analyzer in the 1980's, measurement of material properties in waveguide or coaxial test fixtures combined with other low-frequency analyzer data, the coverage from kHz to 20 GHz can be easily achieved.[4] However, the application of KK relations to deduce one constitutive parameter from another is a trivial question in the radio frequency, microwave, and millimeter wave bands. In such case, the measurements of complex reflection and transmission coefficients provide a direct mean (by proper modeling) of calculating permittivity and permeability. Hence, the KK analysis is mostly applied to validate causality, improve data continuity, smoothing, or supplementing network analyzer measurements.[4,12] Fig. 3 shows an example of the KK relations used for evaluating the quality of the experimental complex permeability data of MnZn ferrite.[12] There are no significant differences between the measured results (solid lines) and KK analysis of $\mu'(f)$ and $\mu''(f)$, and thus it is assumed that the data can be used for further analysis.

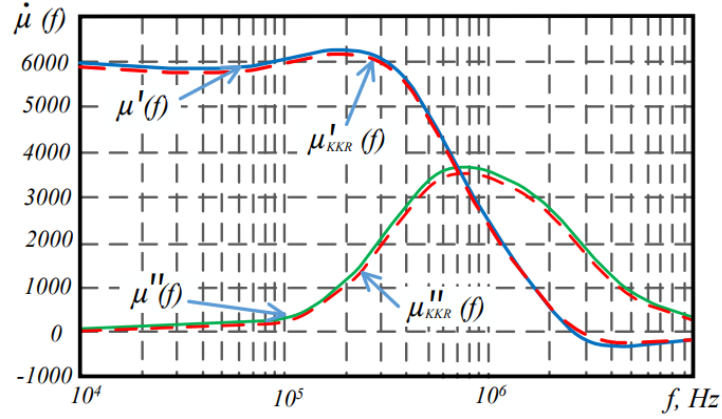

FIG. 3. Application of KK relations to experimental data of complex permeability of MnZn ferrite sample. Here $f=\omega/2\pi$ (reprinted with permission from Ref.12, copyright 2016 De Gruyter Company).

**C. Standard models for frequency dispersion of EM constitutive parameters**

In the frequency domain, dielectric and magnetic loss mechanisms mainly correspond to a relaxation or resonant type. With respect to the microwave frequency range, dielectric losses are related to relaxation behavior,[11] while magnetic losses relate to resonant behavior.[4] With the aim of elucidating the dispersion characteristics of EM parameters, various models have been developed.

*1. Debye relaxation model*

Here, we consider simple cases of relaxation phenomena by considering that the driving field is suddenly switched off after being kept constant for a sufficiently long time. After switching off, a random dipole distribution is expected with dipole moment and polarization going to zero. However, this process is not instantaneous because changing the dipolar orientation occurs by interaction or collision with other dipoles. A characteristic time or relaxation time $\tau$ exists (roughly the time between collisions) before the dipole moment vanishes. First consider the Debye model for which the dipoles are characterized by a single relaxation time. The Debye formulation in the frequency domain of the real and imaginary parts of the permittivity is of the form[2]

$$\varepsilon_r(\omega)=\varepsilon_\infty+\frac{\varepsilon_s-\varepsilon_\infty}{1+(j\omega\tau)}, \qquad (17)$$

$$\varepsilon_r'(\omega) = \varepsilon_\infty + \frac{\varepsilon_s - \varepsilon_\infty}{1+\omega^2\tau^2} \quad , \quad \varepsilon_r''(\omega) = \frac{\omega\tau(\varepsilon_s - \varepsilon_\infty)}{1+\omega^2\tau^2}. \tag{18}$$

The real part of dielectric permittivity $\varepsilon'$ varies between a maximum value $\varepsilon_s$, at a static frequency (i.e., the frequency just before the dipolar relaxation) and a minimum value $\varepsilon_\infty$, at very high frequencies (but not high enough to involve any resonant processes of electronic and atomic polarization). The imaginary part $\varepsilon''$ is related to losses and is maximal at angular frequency $\omega = 1/\tau$ (Fig. 4a). The Debye relaxation model characterizes well the dielectric behavior of many gaseous and some liquid materials with dipolar molecules.[13] However, for most solids, the loss peak becomes much broader and cannot be expressed by a single relaxation time $\tau$ but is merely represented by a distribution of relaxation times.[14] In addition, the dipoles are more likely to be interactive and their response depends of the alternating field. Since the Debye model cannot properly predict the dielectric behavior of many materials, several relaxation functions have been heuristically derived from the Debye equation. These expressions include added parameters which, for specific limiting conditions, reduce to the Debye equation. For example, Cole and Cole equation [15] is expressed by

$$\varepsilon_r(\omega) = \varepsilon_\infty + \frac{\varepsilon_s - \varepsilon_\infty}{1+(j\omega\tau)^{1-\alpha}} \quad 0 \leq \alpha \leq 1, \tag{19}$$

where $\alpha$ is an empirical exponent, corresponding to a measure of the loss peak broadening. For $\alpha=0$, Eq. (19) yields Debye equation. The real and imaginary parts of the relative permittivity are

$$\varepsilon_r'(\omega) = \varepsilon_\infty + (\varepsilon_s - \varepsilon_\infty)\frac{1+(\omega\tau)^{1-\alpha}\sin\left(\frac{1}{2}\alpha\pi\right)}{1+2(\omega\tau)^{1-\alpha}\sin\left(\frac{1}{2}\alpha\pi\right)+(\omega\tau)^{2(1-\alpha)}},$$

$$\varepsilon_r''(\omega) = (\varepsilon_s - \varepsilon_\infty)\frac{(\omega\tau)^{1-\alpha}\cos\left(\frac{1}{2}\alpha\pi\right)}{1+2(\omega\tau)^{1-\alpha}\sin\left(\frac{1}{2}\alpha\pi\right)+(\omega\tau)^{2(1-\alpha)}}. \tag{20}$$

The complex plane (Cole-Cole) plot, i.e. $\varepsilon'$ against $\varepsilon''$, can be used for the analysis of dielectric relaxation. The loss factor is plotted versus the real permittivity, tracing a semicircle if the dielectric relaxation exhibits a single relaxation time, i.e. Debye type relaxation (Fig. 4b). A deviation from a semicircle indicates a distribution of relaxation times or Cole-Cole relaxation. Beyond the Cole-Cole model, there are several other proposed equations such asCole-Davidson[16,17] and Havriliak-Negami[18,19]. The latter is the more general one

$$\varepsilon_r(\omega) = \varepsilon_\infty + \frac{\varepsilon_s - \varepsilon_\infty}{(1+(j\omega\tau)^\alpha)^\beta} \quad 0 < \alpha \leq 1 \quad 0 < \beta < 1, \tag{21}$$

where $\beta$ controlled the asymmetry of the loss peak. When $\beta=1$ the model results in the Cole-Cole model; $\alpha=1$ generates the Cole-Davidson model. It has been also shown that this equation can accurately describe the dynamic mechanical behavior of polymers.[20]

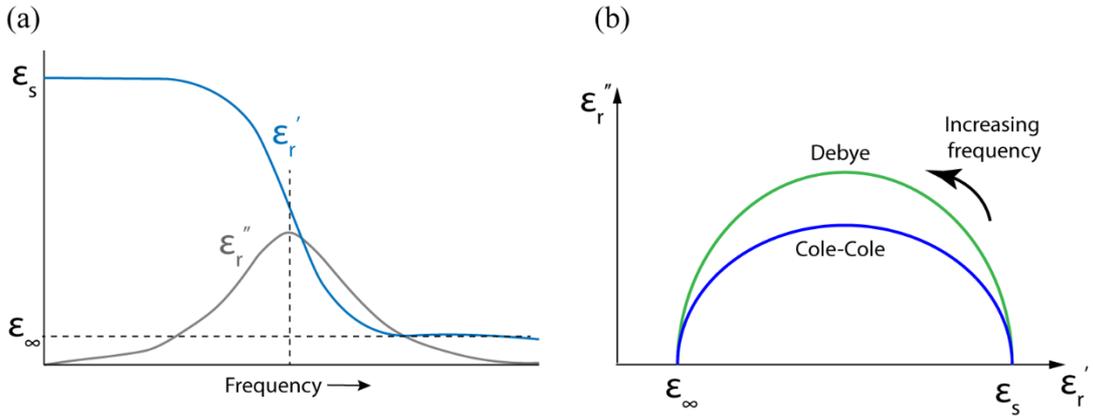

FIG. 4. (a) Sketch of the frequency dependence of the real and imaginary parts of permittivity for a dielectric material exhibiting a Debye relaxation; (b) Cole-Cole plots of Debye and Cole-Cole equations. A semicircle with its center on the real axis is shown in the case of Debye model, while an arc of a semicircle with its center below the real axis corresponds to Cole-Cole model; the apex of the arch is the relaxation frequency.

*2. Lorentz dispersion model*

The Lorentz oscillator model, also known as Drude-Lorentz oscillator model, is based on modeling an electron as a driven damped harmonic oscillator. The binding between electron and nucleus can be described by a hypothetical spring with a specific spring constant and driving force corresponding to the oscillating electric field.[21] A damping force exists so that oscillations do not tend to be infinite when the driving force is at resonance. Lorentz model describes inter-band transitions, i.e. transitions for which electron moves to a final state related to a different band without changing its **k**-vector in the first Brillouin zone.[22] We mention that the Drude-Lorentz model is based on Newton's second law of motion to describe the electron motion allowing us to derive the expressions for the dipole moment, polarization, susceptibility, and permittivity. The real and imaginary parts of the relative permittivity for a material with a single resonance are

$$\varepsilon_r(\omega) = 1 + \frac{\omega_p^2}{\omega_0^2 - \omega^2 - j\omega\Gamma_0}, \quad \omega_p^2 = \frac{Nq^2}{\varepsilon_0 m_e}, \tag{22}$$

$$\varepsilon_r'(\omega) = 1 + \frac{\omega_p^2(\omega_0^2 - \omega^2)}{(\omega_0^2 - \omega^2)^2 + \omega^2\Gamma_0^2}, \quad \varepsilon_r''(\omega) = \frac{\omega_p^2 \omega \Gamma_0}{(\omega_0^2 - \omega^2)^2 + \omega^2\Gamma_0^2}, \tag{23}$$

where $\omega_p$ corresponds to the plasma frequency which depends on the electron density $N$, electron charge $q$ and electron mass $m_e$; $\omega_0$ is the resonance frequency of the oscillator; $\Gamma_0$ is the damping factor (full width at half maximum (FWHM) of the loss peak). As the damping decreases, the peak gets narrower and taller (Fig. 5a).

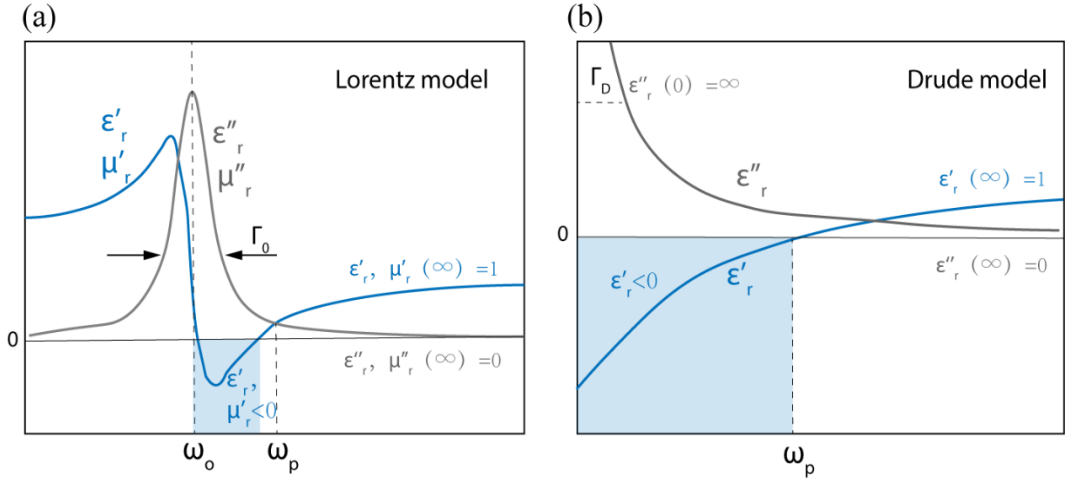

FIG. 5. Graph of the frequency dependence of the real and imaginary parts of permittivity (permeability) for a material exhibiting (a) Lorentz model and (b) Drude model.

The Lorentz model for magnetic permeability is similar to the Lorentz model of electric permittivity except that instead of dealing with the electric field, the driving force in this case is the magnetic field. Therefore[23]

$$\mu_r(\omega) = 1 + \frac{\omega_{mp}^2}{\omega_{0m}^2 - \omega^2 - j\omega\Gamma_m}, \tag{24}$$

where $\omega_{mp}$, $\omega_{0m}$, and $\Gamma_m$ are the plasma frequency, resonant frequency, and damping coefficient in the magnetization of a magnetic dipole. The general form of the real and imaginary parts of the relative magnetic permeability are

$$\mu_r'(\omega) = 1 + \frac{\omega_{mp}^2(\omega_{0m}^2 - \omega^2)}{(\omega_{0m}^2 - \omega^2)^2 + \omega^2\Gamma_m^2}, \quad \mu_r''(\omega) = \frac{\omega_{mp}^2 \omega \Gamma_m}{(\omega_{0m}^2 - \omega^2)^2 + \omega^2\Gamma_m^2}. \tag{25}$$

Fig. 5a shows typical Lorentz response curves for the relative complex permeability in a similar fashion as permittivity.

### *3. Drude dispersion model*

Drude model is based on the classical mechanical theory of free electron and was derived in order to explain the transport properties in metals (due to intra-band transitions), conductive oxides, and heavily doped semiconductors.[24,25] Electrons are free to move, colliding, bouncing, and re-bouncing in relatively static positive ions leading to electrical conduction. Since the conduction electrons are assumed to be free, Drude oscillator is an extension of the Lorentz oscillator model taking the restoring force and resonance frequency as null values ($\Gamma_0 = 0$, $\omega_0 = 0$). Drude's permittivity reads

$$\varepsilon_r(\omega) = 1 - \frac{\omega_p^2}{\omega^2 + j\omega\Gamma_D}, \tag{26}$$

where $\Gamma_D$ is the damping parameter attributed to the delocalized electrons (intra-band transitions) also known as the electron-electron collision frequency. As $\Gamma_D$ increases, the broadening of the loss tail increases as well (Fig. 5b). The real and imaginary parts of the relative permittivity are

$$\varepsilon_r'(\omega) = 1 - \frac{\omega_p^2}{\omega^2 + \Gamma_D^2}, \quad \varepsilon_r''(\omega) = \frac{\omega_p^2 \Gamma_D}{\omega(\omega^2 + \Gamma_D^2)}. \tag{27}$$

Below the plasma frequency $\omega < \omega_p$, $\varepsilon_r'$ is negative (Fig. 5b). In this case, the electric field cannot penetrate the material and thus an incident wave of frequency less than the plasma frequency will not propagate into the material medium and will be totally reflected. When $\omega = \omega_p$, $\varepsilon_r'$ is zero, i.e. all electrons oscillate in phase throughout the material propagation length. At high frequencies above the plasma frequency $\omega > \omega_p$, the reflectivity decreases and the metal becomes transparent to EM waves.

### D. Coupling of permittivity and permeability

Free charge carriers are not only fundamental sources of permittivity but also magnetic permeability. Current source or moving charges generate magnetic field and hence magnetization, and magnetic permeability are expected to be different from the free space value. The current induces a magnetic field with a magnitude proportional to the current and opposed to the external magnetic field.[4] In addition, due to the induced field opposing the external field, it supplies a negative contribution to permeability, and thus when added to the unit relative magnetic permeability the sum is less than unity.[4] Such reduction generates diamagnetism which is rather minor but it is present in all materials with moving charges when there is an external field source. Diamagnetism is especially prominent in composites that contain conducting particulates, even in those containing electrically isolated ferromagnetic particles such as Fe, Ni, or Co. Such response often appears at frequencies ranging from hundreds of MHz to tens of GHz and can be shifted by external DC magnetic fields. Composites combining dielectric and conducting or semiconducting materials display Lorentzian-like dispersions exhibited by magnetic materials which can obtain broader bandwidth responses if the mixture is precisely adjusted. The volume concentration, particle size, and shape control interfacial polarization, and the finite distance over which charge may flow in the composite, is defined as percolation.[26] Percolation is related to the establishment of characteristic lengths and paths that allow charge transport through networks of conducting particles in contact with each other.[4,26] Such networks modify the permittivity frequency dispersion and thus offer a means to manipulate magnetic dispersions. The coupling of frequency dispersive permittivity and permeability derived from an electromagnetic scattering model of particulates due to their finite electrical size, diamagnetism, and shape was first described by Lewin.[27] The Lewin function $F$ couples both permittivity $\varepsilon_1$ and permeability $\mu_1$ of the composite of particulate of radius $r_0$ [27,28]

$$\varepsilon_p = \varepsilon_1 F(\phi), \mu_p = \mu_1 F(\phi), \quad F(\phi) = \frac{2(\sin\phi - \phi\cos\phi)}{(\phi^2 - 1)\sin\phi + \phi\cos\phi}, \quad \phi = 2\pi r_0 \sqrt{\varepsilon_1 \mu_1}/\lambda_0, \quad (28)$$

where $\lambda_0$ is the free-space wavelength.

Researchers have also found that in magneto-electric, ferroic, and chiral materials the application of magnetic field produces a dielectric response and the application of an electric field generates a magnetic response.[29] Such cross-coupling behavior originates from the strain-induced distortion of the spin lattice upon the application of an electric field. For example, in materials such as chromium oxide, the lattice is slightly distorted when applying a strong electric field, changing the magnetic moment and thus the magnetic response.[30] Extrinsic effects can be produced by e.g. specific layering of magnetic, ferroelectric, and dielectric materials in a way that an external electric field modifies the magnetic response and a magnetic field modifies the electric response. Chiral materials can be assembled by embedding conducting spirals into a dielectric matrix. For artificial magneto-electric materials, the permittivity and permeability are effective parameters rather than intrinsic properties. In this case, the constitutive relations for displacement fields are more complex and contain cross coupling between fields.[30]

## III. EFFECTIVE MEDIUM THEORIES AND NUMERICAL MODELS OF ELECTROMAGNETIC COMPOSITES

Accurate, simultaneous control of multiple degrees of freedom is crucial for the engineering of new types of multifunctional composites. For this reason, there has been considerable interest in control techniques to improve electromagnetic wave transport in composites.[31-35] On the coarse-grained modelling side, the commonly designated as effective medium theories have played a pivotal role in defining the electromagnetic parameters, i.e. $\varepsilon$ and $\mu$. Knowing $\varepsilon$ and $\mu$ of composites is key to understanding their polarization and magnetization behaviors and global properties. This important question of fundamental physics also bears critical consequences in technological situations of diverse sorts, e.g. microwave absorption properties of plasto-ferrites, and microwave and millimeter wave devices (tunable phase shifters, resonators, and delay lines) which exploit the magnetoelectric coupling.

A common goal in materials science is the determination of relationships between the microscopic, mesoscopic, and macroscopic scales of a material and its (electrical, magnetic, optical, thermal, mechanical, and so forth) properties. Moreover, control is all-important. It is easy to produce interesting structures by chance, but much harder to build or replicate a chosen structure by design. So it makes sense to look for general principles that can guide construction even when the requirements become complex. Knowledge of such structure-property relationships is crucial for proper selection of engineering materials that meet a required set of specifications. This is a problem of formidable difficulty, and the general results are not known except for very particular, idealized cases. Precisely, because the dependence of macroscopic properties on the internal length scales is not yet completely understood, and precisely because the explanation may lie in any of the areas of materials science, the attack on the problem demands a wide expansion in our understanding of the fundamental processes of computational electromagnetics. However, various approximate methods can be attempted and then compared in order to work out a progressively more detailed and satisfactory picture. For many centuries, materials were discovered and processed in a largely serendipitous way. Fortunately, we are entering an era in which high-performance computing is coming into its own, allowing true predictive simulations of complex materials to be made from information on their individual constituents and their morphology.[36]

The pace at which the theoretical modelling of the electromagnetic properties of composites is advancing is staggering. In the little less than 90 years since the discovery of the Bruggeman homogenization formalism of particulate composites comprising isotropic randomly distributed components within a host matrix,[37,38] the field of EM composites has entered its own age, sometimes dubbed "the computational age".[36] This success story in the emergence of modern theory of composites is the unprecedented and detailed multiscale and multiphysics of particulate structures that has been garnered as a result of the emergence of new ways to engineer composites with specific multifunctionality. A particularly fertile example that serves as the backdrop for the present section is given by our ever-improving understanding of the effective electromagnetic properties of a variety of two-phase composites, i.e. graded and core-shell composites, metamaterials, magnetoplasmonic composites, which have been approached by experimental and numerical analysis, see. e.g. Refs. [39-41].

**A. Effective permittivity and magnetic permeability of composites: a primer**

The goal of this subsection is to develop a feeling for effective electromagnetic parameters. That is, for the many different ways we can think about composites whether structurally, magnetically, or electrically, we will try to formulate those insights in first-principles terms.

With each passing generation, our understanding of the electromagnetic transport in materials is becoming more and more refined. We have learned a huge amount about the multiscale structures of materials and the atoms that make them up. But what about the specific electric and magnetic properties at these different scales? The resulting microscopic heterogeneity linked to randomness in positions of filler particles, their interaction with the host matrix, etc. causes a spatially and temporally non-uniform response to external response even under macroscopically uniform electromagnetic excitation. In many ways, the development of a census of the important length scales characterizing the internal disorder is an astonishing achievement and has revealed not only that all materials are heterogeneous, but that their macroscopic properties depend largely on the hierarchization of these scales. Though there is still much left to be understood about precisely how composites keep track of their property-structure relationships, in this subsection, we focus on what has been learned thus far about this issue from a quantitative perspective.

There are many organizational principles for providing significant length scales that characterize composites. To get rid of the irrelevant details of the morphology of the composites, a reasonable strategy to be adopted here is to organize these scales along three key scales (Fig. 6), starting with the microscale (atoms and dipoles), followed by the mesoscale (filler particles or phases), and finally the macroscale (sample volume).

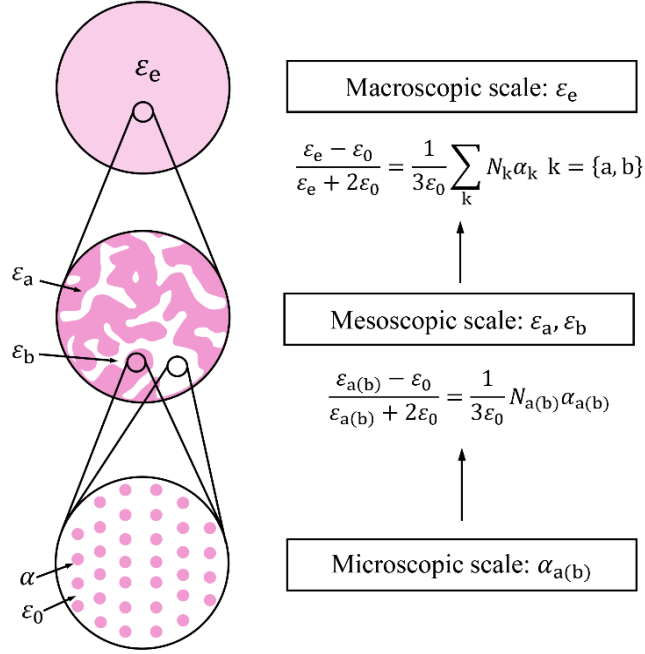

FIG. 6. Diagrams showing qualitatively the three-scale description of a composite that relates microscopic entities (atoms) with those at the mesoscopic scale (phases, inclusions) and then the macroscopic scale (sample volume). The micro-scale (bottom panel) corresponds to the level of interacting entities (atoms and dipoles). The meso-scale (intermediate panel) is relevant to the level of statistical description. The macro-scale (top panel) deals with the sample volume of the composite (adapted from Berthier, *Optique des Milieux Composites*, (Polytechnica, 1993)[35]).

In particular, depending upon the context, there are many different ways of thinking about mesoscale, e.g. filler particle aggregate or agglomerates. Once these scales are in hand, we then attempt to connect all these scales by a multiscale analysis. The advantage of this three-scale procedure is that it is simple enough that the relevant features of the effective permittivity and magnetic permeability of composites can be checked against experimental data.

For our derivation, we assume no physical anisotropies in order to simplify the results and incorporating the rank two tensor notation would unnecessarily obscure and complicate our presentation. We can translate and expand on many of these results in the tensor context to better understand the theory and potential limitations.[33] Although conceptually compelling, we furthermore observe that the language of dielectrics which is used here can be translated for the magnetic composite counterparts without new ingredients in view of the mathematical equivalence of a variety of issues of the linear response theory.[34]

We begin by summarizing important concepts of the dielectrics theory. We defined previously the polarization $\boldsymbol{P}$ as the sum of all multipolar moments which are induced by the electric field. the polarization vector is related to the polarizability of the different constituents of the composites. Restricting us to dipolar interactions, $\boldsymbol{P}$ can be expressed as

$$\boldsymbol{P} = \sum_k N_k \alpha_k (\boldsymbol{E}_l)_k, \tag{29}$$

where $N_k$ denotes the density of dipoles of species k, $\alpha_k$ is the polarizability of dipole k, and $\boldsymbol{E}_l$ is the local field at dipole sites k. From the very definition of the macroscopic effective permittivity of the CM, we have

$$\boldsymbol{P} = (\varepsilon - \varepsilon_0)\boldsymbol{E}, \tag{30}$$

where $\varepsilon_0$ is the permittivity of the continuum host 3D medium (this background permittivity does not need to be that of free space), and $\boldsymbol{E}$ is the macroscopic electric field. Making use of Eq. (29) and Eq. (30), we find

$$(\varepsilon - \varepsilon_0)\boldsymbol{E} = \sum_k N_k \alpha_k (\boldsymbol{E}_l)_k. \tag{31}$$

A couple of remarks are in order. On the one hand, Eq. (31) allows us to define a transition between two different levels of description. On the other hand, by considering Eq. (31), we also explicitly assume that: (i) the dipolar approximation is valid as well as for atoms and inclusions, and (ii) the electric field is uniform at the dipole (long-wavelength approximation). Hence, by connecting the meso-macro scales transition we obtain the permittivity of the effective medium

$$(\varepsilon_e - \varepsilon_0) \boldsymbol{E} = \sum_k N_k \alpha_k (\boldsymbol{E}_m)_k, \tag{32}$$

where $N_k$ denotes the density of inclusions of species k, $\alpha_k$ is the polarizability of dipole k and $\boldsymbol{E}_m$ is the mesoscopic field seen by the inclusion k. In like fashion, the meso-micro scales transition yields

$$(\varepsilon_k - \varepsilon_0) \boldsymbol{E} = N_{ak} \alpha_k (\boldsymbol{E}_l)_k, \tag{33}$$

where $N_{ak}$ denotes the density of atoms of species k. Combining Eq. (32) with Eq. (33) leads to

$$(\varepsilon_e - \varepsilon_0) \boldsymbol{E} = \sum_k \phi_k (\varepsilon_k - \varepsilon_0)(\boldsymbol{E}_m)_k / (\boldsymbol{E}_l)_k, \tag{34}$$

where we introduce the volume fraction of component material k in the CM as $\phi_k$. There are many analytical methods to estimate $(\boldsymbol{E}_m)_k / (\boldsymbol{E}_l)_k$, e.g. it is easy to write down this ratio in the Lorentz approximation of the local electric field and show that it reads as $\boldsymbol{E}_l = \frac{\varepsilon + 2\varepsilon_0}{3\varepsilon_0} \boldsymbol{E}$.[8] Therefore, on account of Lorentz approximation and going back to Eq. (31), we can also write

$$(\varepsilon - \varepsilon_0)/(\varepsilon + 2\varepsilon_0) = \frac{1}{3\varepsilon_0} \sum_k N_k \alpha_k, \tag{35}$$

This gives precisely the Clausius-Mossotti relation which relates the permittivity to the polarizability for material structures for which the Lorentz local field approximation holds.[33-35] The meaning of Eq. (35) needs discussion. Firstly, it allows us to calculate the permittivity for a large variety of materials because we know how to calculate the polarizability. Secondly, from this equation we can estimate the polarizability by physically measuring the permittivity.

Returning to the transition between the mesoscopic (inclusion) to macroscopic (volume) descriptions, we see that

$$(\varepsilon_e - \varepsilon_0)/(\varepsilon_e + 2\varepsilon_0) = \frac{1}{3\varepsilon_0} \sum_k N_k \alpha_k, \tag{36}$$

which then reduces to

$$(\varepsilon_e - \varepsilon_0)/(\varepsilon_e + 2\varepsilon_0) = \sum_k \phi_k (\varepsilon_k - \varepsilon_0)/(\varepsilon_k + 2\varepsilon_0). \tag{37}$$

As a consequence of identifying $\varepsilon_0 = \varepsilon_m$, i.e. one of the components defines a continuum medium in which the inclusions are embedded (Fig. 6), the above equation reduces to the familiar form of Maxwell Garnett (MG) form

$$(\varepsilon_e - \varepsilon_m)/(\varepsilon_e + 2\varepsilon_m) = \sum_k \phi_k (\varepsilon_k - \varepsilon_m)/(\varepsilon_k + 2\varepsilon_m). \tag{38}$$

This should be contrasted if we set $\varepsilon_0 = \varepsilon_e$, i.e. every inclusion is embedded in the same average (effective) medium. In this case, the effective permittivity takes a different form

$$0 = \sum_k \phi_k (\varepsilon_k - \varepsilon_e)/(\varepsilon_k + 2\varepsilon_e), \tag{39}$$

which is the symmetric Bruggeman equation.[8,33-36]

**B. Overview of effective medium theories and their limitations**

As a pedagogical device, we specialize our discussion to a simple example of the homogenization of two isotropic component materials (labeled as 1 and 2) of a particulate continuum composite material. The two-component materials are characterized by the permittivity (or, equivalently magnetic permeability) scalars $\varepsilon_1$ and $\varepsilon_2$. Component material 2 is the randomly dispersed phase. No particular topology is assigned to component particles 2, unless specified. The volume fractions of component materials 1 and 2 are $\phi_1$ and $\phi_2$, respectively, with $\phi_1 + \phi_2 = 1$. Generalization to an arbitrary number of phases is made explicitly by making use of Eq. (39). Each component k (=1,2) is assumed to have a specific shape, e.g. spherical. As a result, Eq. (39) is shape-dependent. The inclusion k is randomly embedded into a self-consistently determined effective (artificial) medium of permittivity $\varepsilon$. The inclusions are embedded in a 3D continuum. Provided that the largest relevant wavelength of the electromagnetic wave which probes the composite is much longer than the linear dimensions of the particles and the size of the sample, the mixture of component materials 1 and 2 may be regarded as being effectively homogeneous. The constitutive parameters of the resulting homogenized composite material are estimated using (effective medium) homogenization formalism. As the particles are taken to have no preferred orientation, the corresponding effective medium is an isotropic material characterized by the scalar permittivity $\varepsilon$. From Eq. (39), the estimate of $\varepsilon$ yielded by the Bruggeman formalism reads

$$\phi_1(\varepsilon_1-\varepsilon_e)/(\varepsilon_1+2\varepsilon_e)+(1-\phi_1)(\varepsilon_2-\varepsilon_e)/(\varepsilon_2+2\varepsilon_e)=0. \qquad (40)$$

The key insight of the analysis above is that the effective permittivity is invariant with a permutation of the two indices in Eq. (40). From this, it follows that both component materials are treated on the same footing. In other words, each inclusion in the composite is considered to be in the same homogenized environment.

The key point to take away is that all models of the effective medium formalisms are self-consistent approaches in which the relationships between the microstructure and properties of composites are treated by mean-field analysis. Refs. [8,33-36] provide conceptual approaches for various situations of physical interest. In particular, Ref. [34] expands the range of possible multipolar interactions between inclusions by considering Green's function perturbation formalism and deals with the coherent potential approximation and the average-t formalisms. We finally comment briefly on Bruggeman equation for materials that contains anisotropic inclusions, either of shape (like rods and tubes) or physical (semi-crystalline phases). For such materials, the scalar picture leading to Eq. (40) is inadequate and the effective medium analysis requires a calculation that addresses fully the tensor representation of the permittivity, i.e. depolarization tensor.[32,33]

Overall, effective medium methods provide a framework to accurately model the electromagnetic and thermal properties of composites. Hence, we see that wide range of effective medium models of two-phase composites is possible. Of course, there may be serious limitations of effective medium analysis. For example, this kind of mean-field analysis is impractical near the percolation threshold in percolative composites.[32-35] Additionally, the effective medium features of physical quantities are valid only in the limit of long-wavelength limit of the electromagnetic field. For smaller wavelengths, some of the features change and new physical effects become important, e.g. scattering.[32-35] The possibility that one or more of the assumptions and simplifications that are made in the effective medium calculation are affecting the electromagnetic parameter does exist, and limitations of this analysis should be understood in that context. One feature that could be looked at more carefully to understand the electromagnetic properties of metamaterials is the coupling between polarization and magnetization phenomena in the presence of external electric and magnetic fields, giving enough opportunities for manipulating devices.[42,43] It should, perhaps, be emphasized in this context that $\varepsilon$ and $\mu$ are independent of each other in the MG (as well as in the Bruggeman) formalism. For instance, Lamb and coworkers have shown that $\varepsilon\mu = \varepsilon_{MG}\mu_{MG}$ in the quasistatic limit.[44]

Today it is generally accepted that some unknown statistical information related to the randomness of inclusions with properties not described by the standard effective medium theory is primarily responsible for the observed disagreement with a diverse set of experimental data and inconsistencies with the predictions of some numerical simulations. Despite the impressive experimental and theoretical work of many researchers the true nature of effective medium description of a composite which should rely on a statistical theory remains unresolved.[33,34] Over the years, a variety of methodologies for probing effective medium theories have been implemented, several of which are summarized in Refs. [32-34,36]. No particular methodology has been generally accepted as ideal. All of these efforts have their own particular advantages, but the list of questions that they can answer is inherently limited by the assumptions that they make. This is the reason why computational electromagnetics techniques can supplement conventional modeling methods to help provide new insights into our basic knowledge of composites.[36]

## C. Comparing with numerical results

The philosophy of electromagnetics of composites has in broad terms, been one of determining statistically-averaged properties of materials, i.e. relating the macroscopic properties of materials to their structure.[32] Until the past recent decades, the characterization of materials characterized by space-varying parameters, i.e. permittivity, and magnetic permeability, has not been easy. It is disconcerting that even for the simplest model thought to capture the salient features of such systems the exact calculation of the permittivity has not yet been established beyond reasonable doubt. As the sophistication of modern theoretical techniques grows physicists have started to seek an understanding and control of the interaction of electromagnetic waves with actual materials that extends beyond mere heuristic descriptions. The problem has been remedied to some extent by numerical computation. Computer simulation is one of the most active and productive areas of research in condensed matter physics and materials science. We think it is fair to recognize that computational physics is a worthy partner of experiment and theory. This has led to the concept of computer simulation as a "third way" of doing science, closer to experiment than theory but complementary to both. Like experiments, simulations produce data rather than theories and should be judged on the quality of those data. For this reason, a purely theoretical design of new materials is well in the future. Approaches such the ones presented here can make the first step in such a process. By identifying a set of possible candidates satisfying a set of criteria we can make a good initial guess and this in itself can speed up development processes tremendously. The "computational" laboratory may therefore be able to narrow down the number of experiments needed for the development of new functionalized materials. This approach exploits the ever-growing power of computers and computational methods in an area of interdisciplinary research that has become known as computational electromagnetics. Computational electromagnetism is commonly based on a discretization of Maxwell's field equations. A number of numerical techniques are available to deal with this subject matter, for example, the finite-element method and the boundary integral method have proved to be enormously fruitful in the design of materials properties.[36] This approach exploits the ever-growing power of computers and computational methods in an area of interdisciplinary research that has become known as computational electromagnetics.[36] There are several reasons for this fruitfulness. Exact numerical techniques have been difficult to formulate and implement. This limitation is due to the difficulty of applying boundary conditions at these complex media. Moreover, materials properties often cannot be attributed to a single phenomenon. Rather, there these properties result from phenomena occurring at many different scales of length and interfaces.

In light of these debates, it is worthwhile to investigate the construction of a virtual composite from a strictly computational viewpoint to see where at least some of the conclusions based on Eq. (40) might arise.

## IV. MATERIAL SCIENCE PERSPECTIVE OF EM COMPOSITES

From the composites science perspective, the effective physical parameters depend on not only the intrinsic characteristic and volume fraction of components but also on the shape and spatial dispersion of inclusions in the microstructure. Hence, an understanding of the relationship between the shape and spatial dispersion of inclusions and the macroscopic electromagnetic properties is essential for a rational design of composite materials with specific application.

### A. Composites with shaped inclusions

The inclusions are called shaped if two or more of the lateral dimensions of the particles are significantly different such as ellipsoids, fibers, and disk-like or needle-like particles.[45] Numerical calculation is required for analyzing the polarizability of inclusions of complex shapes. However, simple analytical solutions can be found for shapes of ellipsoids. A two-component composite where the isolated ellipsoids with a permittivity $\varepsilon_i$ (inclusion volume fraction $\emptyset_2$) are dispersed in a continuous medium of permittivity $\varepsilon_m$ is studied here to demonstrate the influence of shaped inclusions on the effective permittivity of composites. The discussion of the two-component case can be extended to multiphase composites.

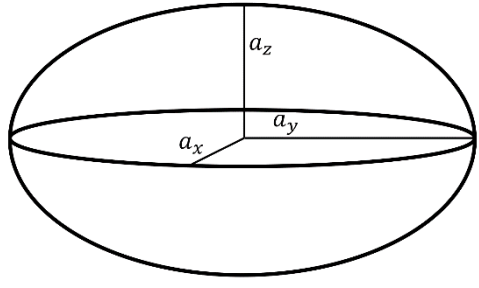

FIG. 7. The geometry of an ellipsoid.

The important parameters in the geometry of an ellipsoid are its depolarization factors. As shown in Fig. 7, the semi-axes of an ellipsoid in the orthogonal directions are $a_x$, $a_y$, and $a_z$. The depolarization factor $N_x$ (the factor in the $a_x$-direction)[46] is

$$N_x = \frac{a_x a_y a_z}{2} \int_0^\infty \frac{ds}{(s+a_x^2)\sqrt{(s+a_x^2)(s+a_y^2)(s+a_z^2)}}. \tag{41}$$

For the other depolarization factor $N_y$ ($N_z$), interchange $a_y$ and $a_x$ ($a_z$ and $a_x$) in Eq. (41). For any ellipsoid, the three depolarization factors satisfy

$$N_x + N_y + N_z = 1. \tag{42}$$

A sphere has three equal depolarization factors of 1/3. The two special cases of ellipsoids are a needle (depolarization factor 0, 1/2, 1/2) and a disc (1, 0, 0). For ellipsoids of revolution, prolate and oblate ellipsoids, Eq. (41) has the closed-form expressions.[47] Prolate spheroids ($a_x > a_y = a_z$) have

$$N_x = \frac{1-e^2}{2e^3}\left(\ln\frac{1+e}{1-e} - 2e\right), \tag{43}$$

$$N_y = N_z = \frac{1}{2}(1 - N_x), \tag{44}$$

where the eccentricity is $e = \sqrt{1 - a_y^2/a_x^2}$. For oblate spheroids ($a_x = a_y > a_z$),

$$N_z = \frac{1+e^2}{e^3}(e - \tan^{-1}e), \tag{45}$$

$$N_x = N_y = \frac{1}{2}(1 - N_z), \tag{46}$$

where $e = \sqrt{a_x^2/a_z^2 - 1}$. Fig. 8 shows the depolarization factors for prolate and oblate spheroids as functions of axis ratio ($a_z/a_x$).

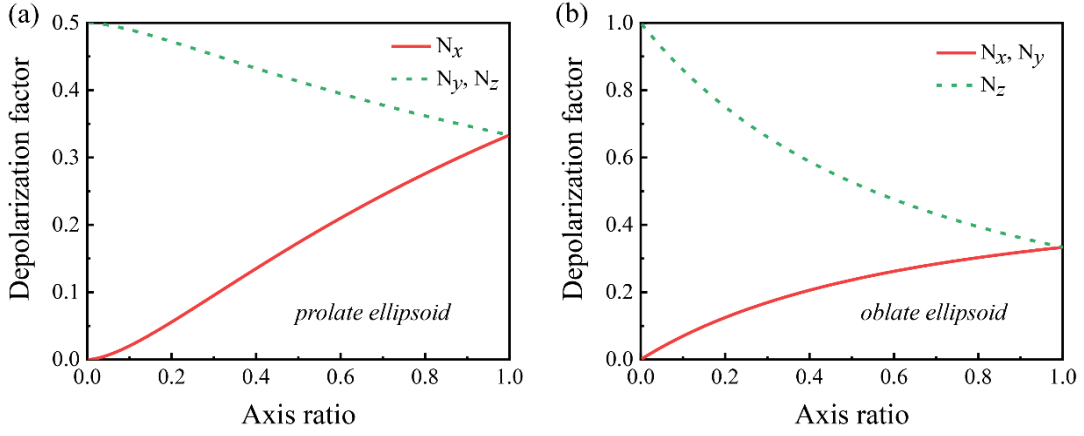

FIG. 8. The depolarization factors of (a) prolate and (b) oblate ellipsoids as functions of the axis ratio $a_z/a_x$.

The polarizability of an ellipsoid is dependent on the direction of the applied electric field because of its geometrical asymmetry. The polarizability component of the ellipsoid for the *x*-directed field is

$$\alpha_x = \frac{4\pi a_x a_y a_z}{3}(\varepsilon_i - \varepsilon_m)\frac{\varepsilon_m}{\varepsilon_m + N_x(\varepsilon_i - \varepsilon_m)}. \tag{47}$$

Replacing $N_x$ with $N_y$ and $N_z$, the *y*-and *z*-direction components of the polarizability can be obtained. Now consider a mixture a composite in which all the ellipsoids are orderly-oriented. The effective permittivity of the composite is anisotropic. The Maxwell Garnett formula for the *x*-component of the effective permittivity of the composite is

$$\varepsilon_{e,x} = \varepsilon_m + \emptyset_2 \varepsilon_m \frac{\varepsilon_i - \varepsilon_m}{\varepsilon_m + (1-\emptyset_2)N_x(\varepsilon_i - \varepsilon_m)}. \tag{48}$$

Similarly, $\varepsilon_{e,y}$ and $\varepsilon_{e,z}$ can be written by replacing $N_x$ by $N_y$ and $N_z$, respectively. At the given inclusion volume fraction $\emptyset_2$, the component of effective permittivity decreases with the increase of the depolarization factor, as shown in Fig. 9(a). Another common case for a composite is that ellipsoids are randomly oriented, which is an isotropic mixture. The effective permittivity $\varepsilon_e$ is a scalar. One-third of each polarizability component gives equal shares to the macroscopic polarization density, and the expression of $\varepsilon_e$ is given by[47]

$$\varepsilon_e = \varepsilon_m + \varepsilon_m \frac{\frac{\phi_2}{3}\sum_{j=x,y,z}\frac{\varepsilon_i-\varepsilon_m}{\varepsilon_m+N_j(\varepsilon_i-\varepsilon_m)}}{1-\frac{\phi_2}{3}\sum_{j=x,y,z}\frac{N_j(\varepsilon_i-\varepsilon_m)}{\varepsilon_m+N_j(\varepsilon_i-\varepsilon_m)}}. \tag{49}$$

Obviously from Fig. 9(b), it can be found that inclusions of discs create a larger effective permittivity than inclusions of needles and spheres. The above cases of orderly- and randomly-dispersion of ellipsoids with different geometries demonstrate that the geometry of inclusions is important for the effective permittivity of composites.

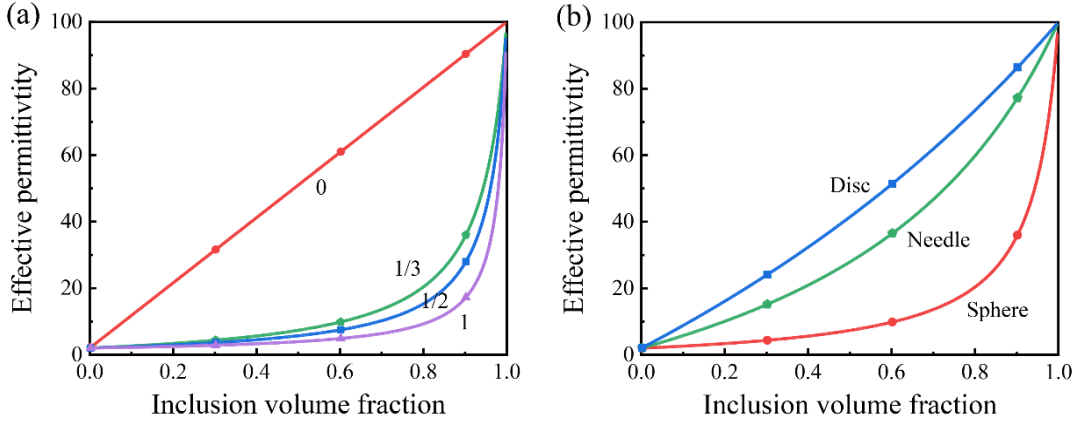

FIG. 9. The effective permittivity of a composite where ellipsoid inclusions of $\varepsilon_i = 100$ are embedded in the matrix of $\varepsilon_m = 2$. (a) The components of the effective permittivity of composites in which aligned ellipsoid inclusions with different depolarization factors of 0, 1/3, 1/2, and 1. (b) The effective permittivity of composites where inclusions (spheres, needles, and discs) are randomly oriented.

### B. Composites with inclusions of different spatial dispersion

Inclusions can be dispersed in the matrix in different spatial architectures. Besides the homogeneous dispersion of inclusions, there are some microstructurally inhomogeneous composites. Fig. 10 illustrates the four types of inhomogeneous dispersion system: (a) inclusions are agglomerated in the form of isolated clusters, (b) inclusions are dispersed in the bar or laminated-like region, (c) 3D network of inclusion-rich phase with isolated inclusion-lean phase, and (d) bi-continuous inclusion-rich phase and inclusion-lean phase.[48] It is complicated or even unsolvable to obtain the exact analytical solution of effective permittivity for these inhomogeneous composites. Therefore, we focus on the bounds rather than accurate solutions for the effective material properties in order to intuitively demonstrate the influence of spatial dispersion on effective permittivity.

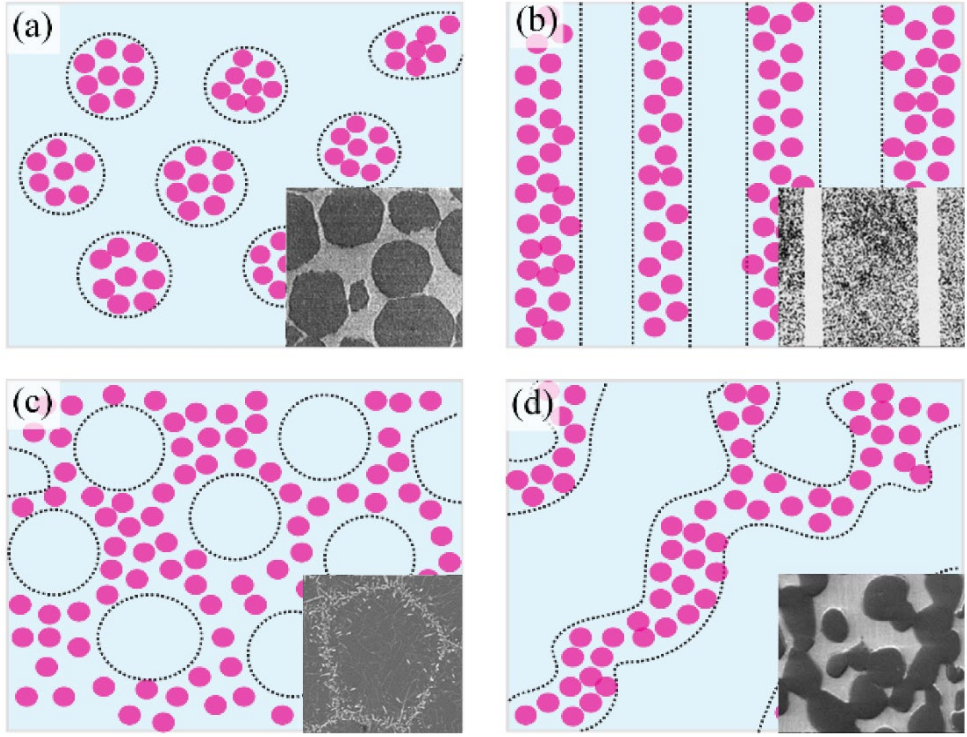

FIG. 10. Schematic illustrations and representative SEM images (insets) of inhomogeneous composites. (a) Patten A: isolated cluster, (b) Patten B: inclusion dispersed in the bar/laminated-like region, (c) Pattern C: 3D network with isolated inclusion-lean phase, and (d) Pattern D: bi-continuous inclusion-rich phase and inclusion-lean phase (reproduced with permission from Ref. 48, copyright 2015 Elsevier).

Let us consider a two-phase composite, where the permittivity of the phases is $\varepsilon_i$ and $\varepsilon_m$, and the volume fraction of $\varepsilon_i$ phase is $\emptyset_2$. It is well known that the classical rule of mixtures (RoM) model (sometimes called Wiener bounds) can be used to estimate the effective property of composites.[48,49] According to the theoretical model, the effective permittivity of all the composites should be located between the RoM upper ($\varepsilon_{\text{RoM,max}}$) and lower ($\varepsilon_{\text{RoM,min}}$) bounds

$$\varepsilon_{\text{RoM,max}} = \emptyset_2 \varepsilon_i + (1 - \emptyset_2)\varepsilon_m, \tag{50}$$

$$\varepsilon_{\text{RoM,min}} = \frac{\varepsilon_i \varepsilon_m}{\emptyset_2 \varepsilon_m + (1-\emptyset_2)\varepsilon_i}. \tag{51}$$

$\varepsilon_{\text{RoM,max}}$ and $\varepsilon_{\text{RoM,min}}$ are the effective permittivity of the composites along the

longitudinal direction and transversal direction, as shown in Fig. 11(a). The RoM upper and lower bounds are usually applied to roughly predict the highest and lowest properties and monitor the experimental data. When the needle-like inclusions are aligned with the needle-axis parallel to the electric field (or when the electric field is tangential to the surface of the disk-like inclusions), the effective permittivity saturates to RoM upper limit (or RoM lower limit, respectively).[45]

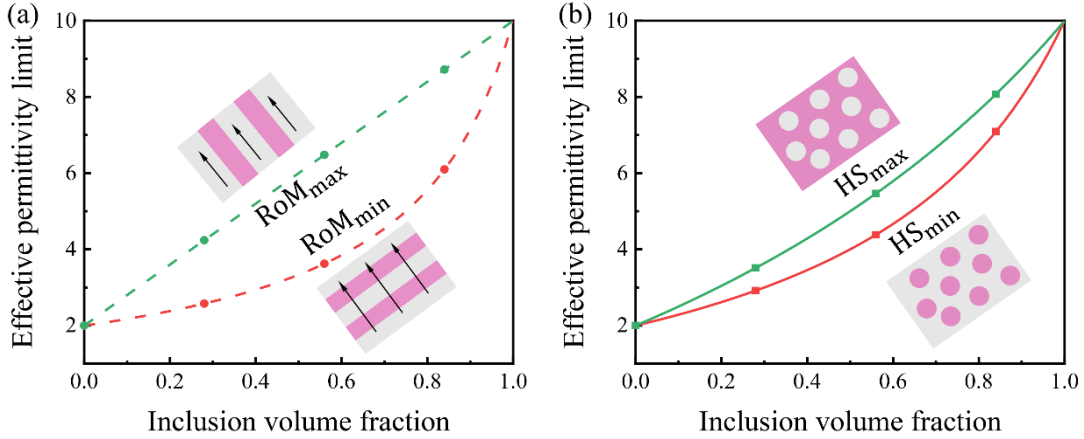

FIG. 11. (a) Rule of mixture bounds and (b) Hashin-Shtrikman bounds for the effective permittivity of a composite with one phase (purple) of $\varepsilon_i = 10$ and another phase (gray) of $\varepsilon_m = 2$.

The composite corresponding to the RoM bounds is anisotropic. The bounds for the macroscopically isotropic composite need to be refined. Hashin and Shtrikman proposed the well-known H-S bounds to effectively estimate the effective permittivity of isotropic composites[50]

$$\varepsilon_{HS,1} = \varepsilon_m + \frac{\emptyset_2}{\frac{1}{\varepsilon_i - \varepsilon_m} + \frac{1 - \emptyset_2}{3\varepsilon_m}}, \quad (52)$$

$$\varepsilon_{HS,2} = \varepsilon_i + \frac{1 - \emptyset_2}{\frac{1}{\varepsilon_m - \varepsilon_i} + \frac{\emptyset_2}{3\varepsilon_i}}, \quad (53)$$

$$\min(\varepsilon_{HS,1}, \varepsilon_{HS,2}) \leq \varepsilon_e \leq \max(\varepsilon_{HS,1}, \varepsilon_{HS,2}). \quad (54)$$

The bounds are equal to the Maxwell Garnett mixing rules for isotropic spherical inclusions. $\varepsilon_{HS,1}$ is the case of spherical inclusions $\varepsilon_i$ in the matrix $\varepsilon_m$, and $\varepsilon_{HS,2}$ corresponds to the case of spherical inclusions $\varepsilon_m$ in the matrix $\varepsilon_i$. The maximum and minimum bounds depend on the relation between the component permittivity. If $\varepsilon_i >$ $\varepsilon_\mathrm{m}$, Eq. (52) is the minimum for the effective permittivity, and Eq. (53) is the maximum. For the case $\varepsilon_\mathrm{i} < \varepsilon_\mathrm{m}$, the situation is the opposite. In the H-S theorem, the upper bound rigorously corresponds to the composite containing the inclusion phase of lower permittivity encapsulated by a continuous matrix phase of higher permittivity, while the lower bound corresponds to the composite with higher-permittivity inclusion phase encapsulated by a lower-permittivity matrix phase, as shown in Fig. 11(b). The H-S bounds are tighter than the RoM bounds. From the analysis of limits for effective permittivity, it can be concluded that the microscopic and macroscopic distribution of phases in composites contribute to the macroscopic effective permittivity of composites.

## V. METAMATERIAL WITH PECULIAR ELECTROMAGNETIC PROPERTIES

### A. Introduction to metamaterials

The permittivity $\varepsilon$ and the permeability $\mu$ are the fundamental characteristic quantities that determine the response of materials and electromagnetic waves. $\varepsilon$ and $\mu$ of naturally occurring materials are not simultaneously less than 0. First introduced by Veselago theoretically in 1968 for a material with simultaneously negative $\varepsilon$ and $\mu$, left-handed material (LHM) possessed many new features such as negative refraction, backward wave propagation, reversed Doppler shift, and backward Cerenkov radiation.[51] Sir John Pendry innovatively proposed the models of the thin wire medium whose permittivity is negative and split ring resonator whose permeability is negative in 1996 and 1999, respectively.[52,53] Soon afterwards, David Smith demonstrated a composite medium, based on a periodic array of split ring resonators and continuous wires, which exhibits a frequency region characteristic of simultaneously negative values of effective permittivity and permeability, and confirmed the negative refraction of the LHM experimentally.[54,55] Metamaterial, first known as the left-handed material, is synthesized by Walser and defined as artificial composites consisting of "meta-atoms" at sub-wavelength scales with exotic electromagnetic properties not readily available in nature.[56] The artificial composite with elements of subwavelength scales whose electric sizes are smaller than the free-space wavelength still can be regarded as a continuous medium that can be expressed with effective constitutive parameters (permittivity and permeability). Metamaterials achieve material performance "beyond" the limitations of conventional composites, opening a door to significantly extend material properties and tailoring independently the permittivity and permeability at will.

According to the various combinations of permittivity and permeability, materials are classified into four quadrants, including conventional materials, double-negative materials, negative-permittivity materials, and negative-permeability materials,[57,58] as shown in Fig. 12. Conventional materials with positive permittivity and permeability allow propagating waves and the electric vector (***E***), magnetic vector (***H***), and wave vector (***k***) correspond to the right-handed rule. Materials with double-negative parameters are transparent and allow wave propagation, but the ***E***, ***H***, and ***k*** vectors follow left-handed nature. Electromagnetic waves decay exponentially in materials with negative permittivity or negative permeability. In this section, we will introduce the design philosophy and construction of single-negative and double-negative metamaterials in detail.

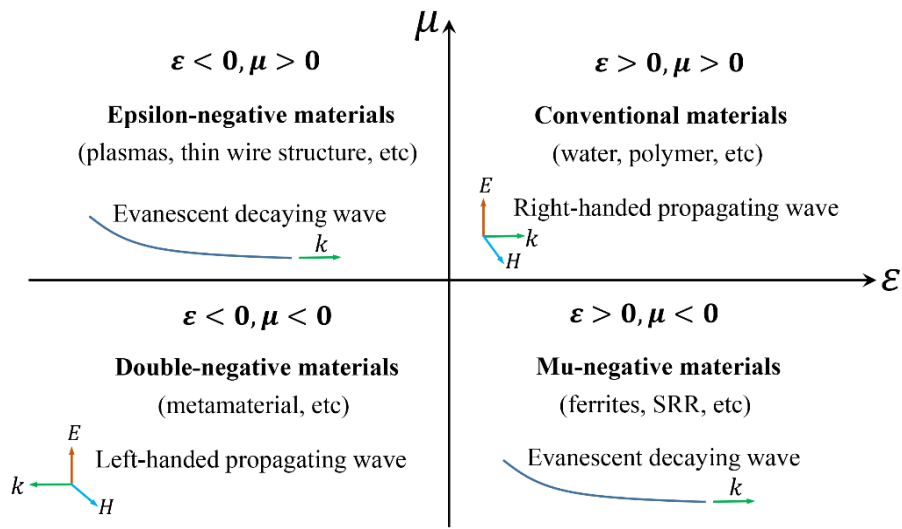

FIG. 12. Categories of materials based on the permittivity and permeability (reprinted with the permission from Ref. 58, copyright 2021 EDP Sciences).

**B. Metamaterials with negative permittivity**

Many metals at optical frequencies have negative dielectric permittivity when the conduction electrons in metals can be assumed to be reasonably free in a background of static positive ion cores. The negative permittivity results from the plasma oscillation of free electrons in the metal under the excitation of electromagnetic waves.[59] The plasma oscillation produces the dielectric flux density **D** opposite to **E**; the permittivity becomes negative. The frequency dispersion of permittivity can be described by Drude model,[52]

$$\varepsilon = 1 - \frac{\omega_p^2}{\omega^2 - j\omega\gamma}, \tag{55}$$

where $\omega_p = \sqrt{n_{eff} e^2 / m_{eff} \varepsilon_0}$ is the plasma frequency, which is determined by the effective electron density ($n_{eff}$) and effective mass ($m_{eff}$) of electrons, and $\gamma$ is the damping factor that represents the dissipation of energy into the system. Generally, the permittivity of metals is negative below the plasma frequency, which is usually in the visible or near-ultraviolet band (e.g. the plasma frequency of aluminum is around $3.62 \times 10^{15}$ Hz). Since the permittivity rapidly decreases to minus infinity with the decreasing frequency, the negative permittivity value is almost minus infinity and the dissipation is giant in the microwave range. The major challenge is to reduce the plasma frequency and extend the plasma behavior to lower frequencies.

*1. Reduced frequency plasma oscillation*

According to the equation of plasma frequency, the reduced plasma frequency can be realized by the low effective electron density and high effective electron mass. Periodic structures of very thin infinitely long metal wires, proposed by Pendry, dilute the average concentration of electrons and considerably enhance the effective electron mass through self-inductance.[52] As shown in Fig. 13(a), very thin metallic wires with a radius of $r$ are assembled into a periodic lattice with a distance $a$. The electrons are confined to move within the thin wires only, reducing the average electron density. The effective electron density ($n_{eff}$) in the structure as a whole is given by

$$n_{eff} = n \frac{\pi r^2}{a^2}, \tag{56}$$

where $n$ is the electron density in these wires. Besides, the thin wires have a large inductance, which acts on the moving electrons in the same manner as mass, and hence the electrons in the wires appeared to be very heavy as Nitrogen atoms.[60] The effective electron mass ($m_{\text{eff}}$) in the wires can be expressed as

$$m_{\text{eff}} = \frac{\mu_0 \pi r^2 e^2 n}{2\pi} \ln(a/r). \tag{57}$$

Therefore, the plasma frequency ($\omega_p$) of the periodic structure is,

$$\omega_p^2 = \frac{2\pi \mu_0 \varepsilon_0}{a^2 \ln(a/r)}. \tag{58}$$

For instance, for aluminum wires, $r = 1\,\mu m$, $a = 5\,\text{mm}$, and $n = 1.806 \times 10^{29}\,\text{m}^{-3}$, $\omega_p$ is around 8.2 GHz. Here the plasma frequency of the array structure is shifted to microwave band that is far below the ones of bulk metals and can be tailored through the macroscopic parameters of the system: wire radius and lattice spacing. That is to say, the performance of metamaterials with negative permittivity originates their properties from their structures rather than directly from the composition.

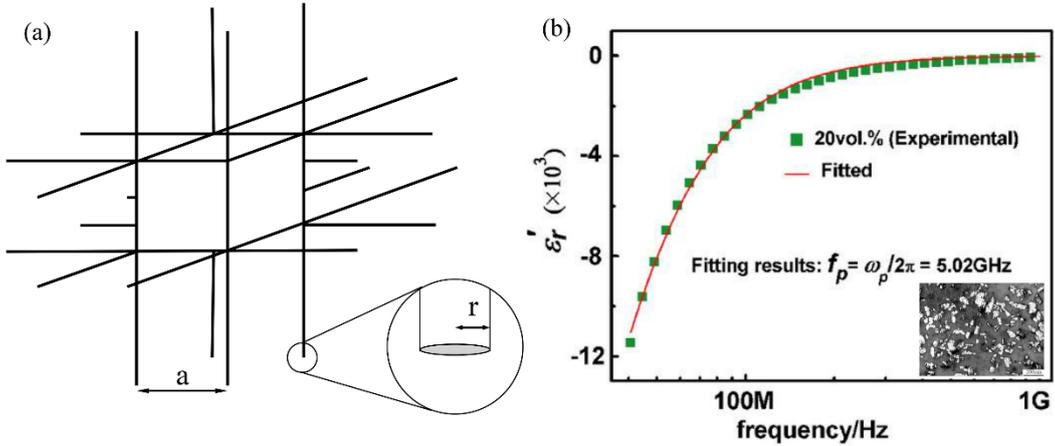

FIG. 13. Negative permittivity arises from reduced frequency plasma oscillation. (a) Schematic of the periodic structure composed of infinite wires arranged in a simple cubic lattice (reproduced with permission from Ref. 52, copyright 1996 American Physical Society). (b) Permittivity spectrum and the optical image for 20 vol.% Fe/Al2O3 composite (reproduced with permission from Ref. 61, copyright 2011 John Wiley and Sons).

Some researchers were dedicated to investigating metamaterials with negative permittivity from a material perspective (composition and microstructure) rather than from an artificial perspective (geometry and architecture). Following the similar idea of electron dilution in thin metallic wires structure, many "real" materials with negative permittivity have been reported and indicated that negative permittivity can be obtained in 3D conductive networks (percolating networks).[58,62,63] For example, the negative permittivity behavior of the $Fe/Al_2O_3$ composite with the content of 20 vol.% at radio frequencies[61] is shown in Fig. 13(b). In conductor/insulator composites, with the conductive filler content exceeding the percolated threshold, the conductive component interconnects and generates conductive networks. In this case, the conductive component dominated the dielectric properties of the whole composites. The composites with conductive networks can be considered as the extensive electrons are diluted by the insolating matrix, hence reducing the effective free electron concentration. The negative permittivity at lower frequencies is attributed to the plasma oscillation of delocalized electrons in the network. Further, the regulation of negative permittivity of the composites can be achieved through composition design and microstructure. For example, graphene/polyolefin composites possess adjustable weak-negative permittivity properties by adjusting the content and reduction methods of graphene.[64]

*2. Electric resonance*

The negative permittivity can be accomplished in the aforementioned plasma-like composites, of which effective permittivity exhibits Drude-type dispersion characteristic. In addition, negative permittivity also appears near a resonance frequency in the Lorentz-type dispersion. We will focus on the negative permittivity properties resulting from electric resonances.

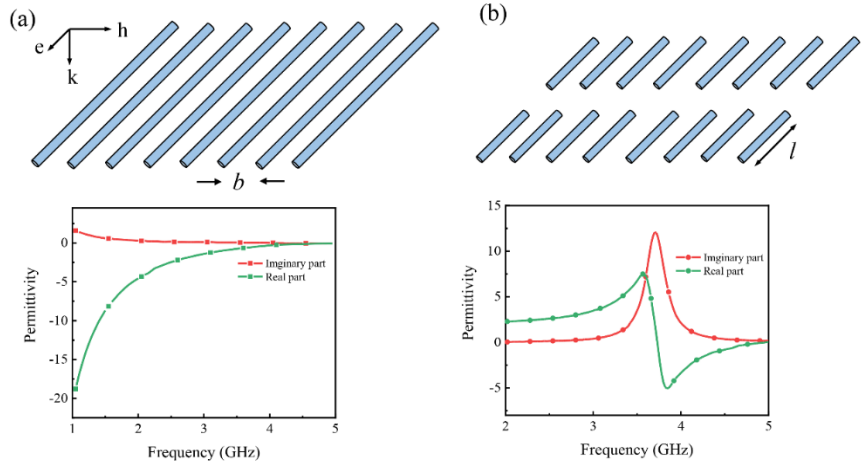

FIG. 14. Sketch and the effective permittivity spectra of the wire composites: (a) long wires and (b) short wires.

*(1). Dipole antenna resonance*

Composites containing long parallel metal wires support very low-frequency plasma oscillation and can be characterized by the Drude-type dispersion of effective permittivity with a negative value of the permittivity below the plasma frequency ($\omega_p$), as shown in Fig. 14(a). How about the short-cut metal wire inclusions? Studies show that the composites with short wire inclusions are characterized by a resonance-type dispersion of the effective permittivity as the wires behave as dipole antennas with the resonance at half wave length conduction: $f_R = c/2l\sqrt{\varepsilon_m}$, where $\varepsilon_m$ is the permittivity of the supporting matrix.[65] For a sufficiently sharp resonance and low wire resistivity, negative dielectric permittivity can be obtained within a frequency band above the resonant frequency, as shown in Fig. 14(b).

*(2). Inductance-capacitance resonance*

The split ring resonators (SRR) are originally utilized to create negative permeability,[53] which will be introduced in detail later. With the development of the SRR structure, some deformed SRRs can exhibit a resonant response to the electric field and support negative permittivity properties, as shown in Fig. 15.[66] The physical nature of the SRR is inductance-capacitance (*LC*) resonance with resonance frequency $\omega = \frac{1}{\sqrt{LC}}$. The electric resonance resulting from the capacitive element couples strongly to the electric field and inductive loops do not.

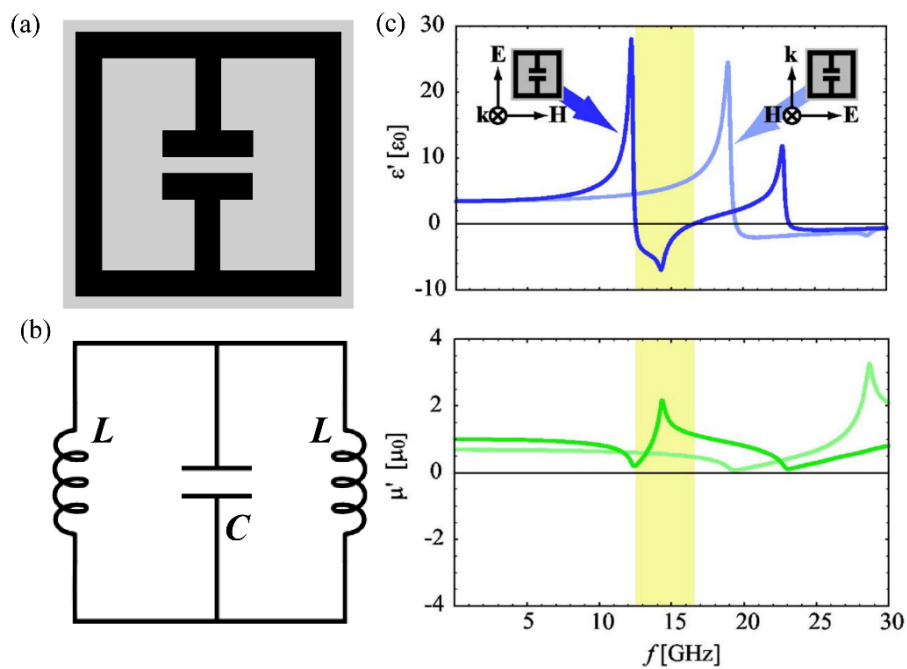

FIG. 15. (a) Schematic, (b) an equivalent circuit, and (c) permittivity and permeability spectra of an electric split ring resonator (reproduced with permission from Ref. 66, copyright 2006 AIP publishing).

*(3). Dielectric resonance*

Besides the above subwavelength metallic structures, dielectric materials can support some resonance modes. The Mie resonance of dielectric particles provides a novel mechanism for the creation of magnetic and electric resonance, hence resulting in negative permittivity and permeability in certain frequency ranges.[67-69] Lewin derived an expression based on Mie scattering theory for the effective permittivity ($\varepsilon_e$) and permeability ($\mu_e$) of the artificial material formed by an array of lossless magnetodielectric spherical particles with permittivity $\varepsilon_2$ and permeability $\mu_2$ embedded in a background matrix with permittivity $\varepsilon_1$ and permeability $\mu_1$, as shown in Fig. 16.[69] The $\varepsilon_e$ and $\mu_e$ of the composite are as follows,

$$\varepsilon_e = \varepsilon_1\left(1 + \frac{3v}{\frac{F(\theta)+2K_e}{F(\theta)-K_e}-v}\right), \tag{59}$$

$$\mu_e = \mu_1\left(1 + \frac{3v}{\frac{F(\theta)+2K_m}{F(\theta)-K_m}-v}\right), \tag{60}$$

where $K_e = \frac{\varepsilon_1}{\varepsilon_2}$, $K_m = \frac{\mu_1}{\mu_2}$, $v = \frac{4\pi a^3}{3p^3}$, $\theta = 2\pi f a\sqrt{\varepsilon_2\mu_2}/c$, $F(\theta) = \frac{2(\sin\theta - \theta\cos\theta)}{(\theta^2-1)\sin\theta + \theta\cos\theta}$, a and p are the sphere radius and the lattice constant, respectively. $F(\theta)$ is a resonant function and becomes infinite at some values of $\theta$, which results in magnetic and electric resonances. When the electromagnetic wave interacts with the subwavelength dielectric particle, displacement current arising from the dielectric polarization can equivalently generate electric dipole or magnetic dipole inside the particle, hence resulting in strong response with the electric or magnetic field, respectively. The effective electric dipoles or magnetic dipoles produce resonance behavior, enabling negative effective permittivity or permeability in a narrow band near the resonance frequency. The constitutive parameters and geometric parameters of the particle will influence the resonance characteristic.

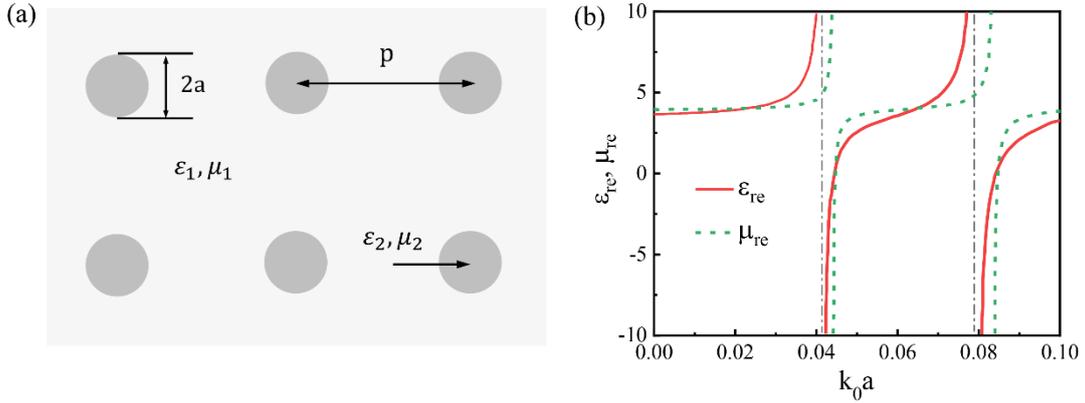

FIG. 16. Schematic (a) and permittivity and permeability spectra (b) of a composite structure containing spherical particles (reprinted with permission from Ref. 69, copyright 2003 IEEE).

Additionally, ferroelectric materials such as $BaTiO_3$ and $BiFeO_3$ presenting intrinsic dielectric resonances at the lower frequencies also can be utilized to accomplish negative permittivity in a specific frequency range.[58,70]

## C. Metamaterials with negative permeability

Magnetic materials have abundant intrinsic magnetic resonances, which can be used to realize negative permeability. Ferromagnetic resonance of magnetic materials under external magnetic field is a good choice to realize negative permeability. For example, Yttrium Iron Garnet (YIG) composite.[59] and ferromagnetic microwires composite.[71] The negative permeability performance can be actively adjusted by varying the external magnetic field. However, the magnetic response of most materials is beginning to tail off at frequencies in the gigahertz range. It is indeed a challenge to obtain magnetic activity, let alone negative magnetic permeability at microwave frequencies and beyond.

Most materials have a natural tendency to be diamagnetic as a consequence of Lentz's law. For an array of conducting cylinders under the illumination of the waves with the magnetic field parallel to the cylinders. The effective permittivity is expressed by[53]

$$\mu_e = 1 - \frac{\pi r^2}{a^2}(1 + i\frac{2\rho}{\omega r \mu_0})^{-1}, \tag{61}$$

where $r$, $a$, and $\rho$ are the radius of cylinder, lattice period, and resistance of the cylinder surface per unit area, respectively. Evidently, $\mu_e$ is never greater than 1 (diamagnetic) or less than 0. The cylinder system showed a limited magnetic effect. From the perspective of equivalent circuit model, the induced currents in the cylinders made the system exhibit an inductive response. By introducing capacitance elements into the system, a rich resonant response can be induced. The array of cylindrical shells with a gap in them (gap width $d$) has been proposed and is well known as the split ring resonator (SRR), where there is a considerable capacitance between the two rings, as shown in Fig. 17(a) and (b). $\mu_e$ with a resonant form and is given by

$$\mu_e = 1 - \frac{\frac{\pi r^2}{a^2}}{1 + \frac{2\rho i}{\omega r \mu_0} - \frac{3 d \mu_0 \varepsilon_0}{\pi^2 \omega^2 r^3}}. \tag{62}$$

$\omega_0$ and $\omega_{mp}$ are defined to be the frequency at which $\mu_e$ diverges and magnetic plasma frequency, respectively

$$\omega_0 = \sqrt{\frac{3 d \mu_0 \varepsilon_0}{\pi^2 r^3}}, \tag{63}$$

$$\omega_{\mathrm{mp}} = \sqrt{\frac{3d\mu_0\varepsilon_0}{\pi^2 r^3(1-\frac{\pi r^2}{a^2})}}. \tag{64}$$

From Fig. 17(c), it can be found that $\mu_e$ is negative when $\omega_0 < \omega < \omega_{\mathrm{mp}}$. Afterward, many deformed SRR structures (e.g. Ω-shaped ring[72] and short wire-pair[73]) were proposed as candidates for negative permeability.

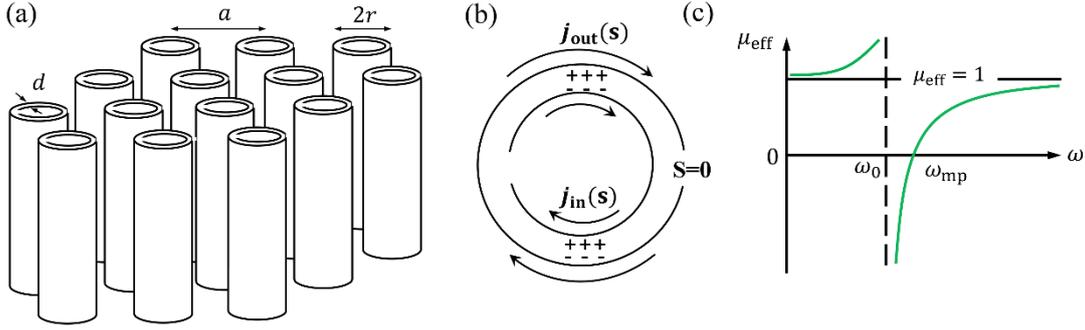

FIG. 17. Negative permeability of Split ring resonators. (a) model of the array of the SRR; (b) current and charge distribution on the SRR; (c) effective permeability of the SRR (reprinted with permission from Ref. 53, copyright 1999 IEEE).

In addition to the inductance-capacitance resonance as discussed above, dielectric resonance such as Mie resonance also can be an alternative to generating negative permittivity medium. The negative effective permeability is attributed to the subwavelength magnetic resonance by the enhancement of the displacement current inside the dielectric "meta-atom".[67]

### D. Metamaterials with simultaneously negative permittivity and permeability

Since we have known how to construct the negative-permittivity and negative-permeability mediums, the combination of both could be considered to realize double-negative metamaterials. Combining a composite with thin wire medium with $\varepsilon < 0$, and the SRR medium with $\mu < 0$ was first proposed and experimentally verified by Smith et al to construct a metamaterial with negative refractive index ($n < 0$).[54] Double negative property can be achieved in the Cu/YIG composites under external magnetic field, where the negative permittivity arises from the percolated metal network and the negative permeability results from the ferromagnetic resonance.[59] Composites containing arrays of high conductive ferromagnetic microwires possess the double-negative-index characteristics for the ferromagnetic resonance and plasma oscillation.[71]

## VI. ELECTROMAGNETIC MEASUREMENT AND APPLICATIONS OF EM COMPOSITES

### A. Electromagnetic measurement techniques

The electromagnetic measurement of composites is required for both scientific and industrial applications. This section describes electromagnetic measurement systems that can be used in microwave frequencies to measure electromagnetic parameters of composites. The measurement systems can be generally divided into two groups, transmission line, and resonant techniques,[4] as shown in Fig. 18. Transmission line techniques are applicable for the measurement of electromagnetic properties of a material over a frequency range, whereas resonant methods are used to attain properties of a material at a specific frequency. We herein focus on the transmission line techniques for the broadband measurement of conventional materials and metamaterials. Significantly, any sample inhomogeneity size scale should be much smaller than both the physical dimension of the sample and electromagnetic wavelength in the sample for the acquirement of effective constitutive parameters.

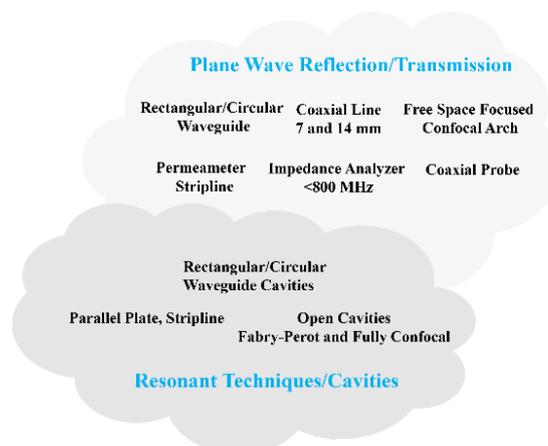

FIG. 18. A cloud of measurement systems (adapted from R. Moore *Electromagnetic*

*Composites Handbook: Models, Measurement, and Characterization*, (McGraw-Hill Education, 2016)[4]).

Transmission line techniques give material constitutive parameters via the reflection coefficient and/or transmission coefficient measured using the network analyzer over the specific frequency range corresponding to the test system.[74] Waveguide system, a common transmission line technique, is sketched in Fig. 19(a). The waveguide system consists of two coaxial-to-waveguide adaptors connected to a rectangular waveguide loaded with the slab of the material under test. The waveguide system is then connected to the network analyzer exciting the incident wave, receiving and processing the reflected and transmitted wave, which constitutes the whole measurement setup. It is noted that the sample should match and completely fill the cross section of the waveguide. The free-space method is also a universal technique due to its non-invasive nature and wideband characterization. An example of a free-space fixture consisting of two antennas and a means of supporting a sample between them is shown in Fig. 19(b).

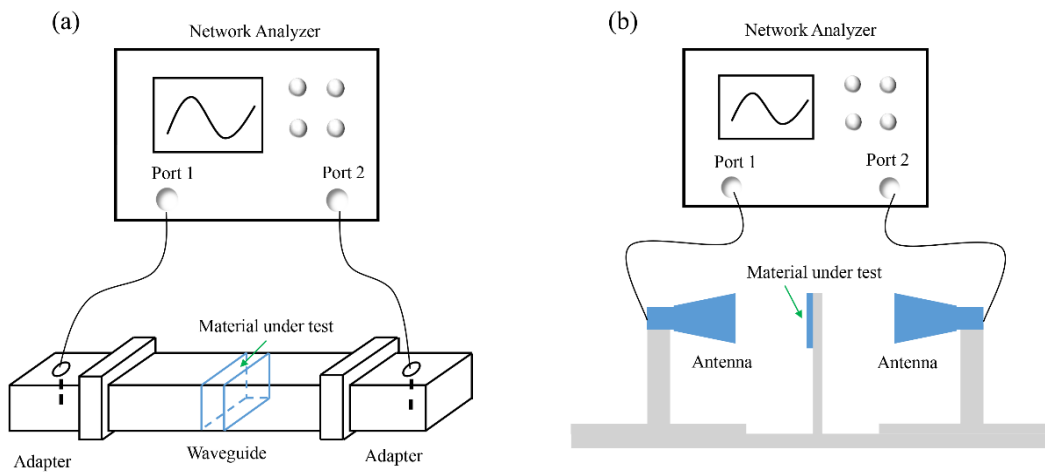

FIG. 19. Experimental scheme of the (a) waveguide and (b) free-space system.

## B. Retrieval method of effective electromagnetic constitutive parameters

### 1. *Nicolson-Ross-Weir method*

Complex permittivity and permeability of materials are determined from the reflection and transmission coefficients. Nicolson-Ross-Weir (NRW) technique is the most commonly used technique for composites, providing direct calculation of both permittivity and permeability from the measured S-parameters.[75,76] The NRW method presumes the sample to be isotropic and homogenous, which can be well satisfied for materials whose inhomogeneity size scale is many times smaller than the electromagnetic wavelength in the material. The retrieval procedure is deduced from the following set of equations.

The reflection ($\Gamma$) and transmission ($T$) coefficients are obtained from the measured Scattering parameters ($S_{11}$ and $S_{21}$)

$$S_{11} = \frac{(1-T^2)\Gamma}{1-\Gamma^2 T^2}, \tag{65}$$

$$S_{21} = \frac{(1-\Gamma^2)T}{1-\Gamma^2 T^2}. \tag{66}$$

Once the S-parameters are extracted from the network analyzer, the reflection coefficient can be given as

$$\Gamma = X \pm \sqrt{X^2 - 1} \ (|\Gamma| < 1), \tag{67}$$

where

$$X = \frac{S_{11}^2 - S_{21}^2 + 1}{2S_{11}}. \tag{68}$$

Therefore, the transmission coefficient can be written as

$$T = \frac{S_{11} + S_{21} - \Gamma}{1 - (S_{11} + S_{21})\Gamma}. \tag{69}$$

The permeability and permittivity can be determined from $\Gamma$ and $T$

$$\mu_r = \frac{1+\Gamma}{\Lambda(1-\Gamma)\sqrt{\frac{1}{\lambda_0^2} - \frac{1}{\lambda_c^2}}}, \tag{70}$$

$$\varepsilon_r = \frac{\lambda_0^2}{\mu_r}\left(\frac{1}{\lambda_c^2} - [\frac{1}{2\pi l}\ln(\frac{1}{T})]^2\right), \tag{71}$$

$$\frac{1}{\Lambda^2} = \left(\frac{\varepsilon_r \mu_r}{\lambda_0^2} - \frac{1}{\lambda_c^2}\right) = -[\frac{1}{2\pi l}\ln(\frac{1}{T})]^2, \tag{72}$$

where $\lambda_0$ is the free-space wavelength, $\lambda_c$ is the cutoff wavelength of the transmission line, and $l$ is the sample thickness. Eq. (72) has an infinite number of roots since the imaginary part of a complex quantity is equal to the angle of the complex value plus $2\pi n$, where $n$ is equal to the integer of $l/\lambda_g$ ($\lambda_g$ is the transmission line guide wavelength). This raises the issue of phase ambiguity, which can be resolved by that the delay through the material is strictly a function of the total length of the material. The calculated group delay for the $n$th solution is

$$\tau_{gn} = l\frac{d}{df}[(\frac{\varepsilon_r \mu_r}{\lambda_0^2} - \frac{1}{\lambda_c^2})_n^{1/2}]. \tag{73}$$

The measured group delay is determined from the slope of the phase ($\phi$) of the transmission coefficient versus frequency

$$\tau_g = \frac{1}{2\pi}\frac{d(-\phi)}{df}. \tag{74}$$

The correct root ($n=k$) is found when $\tau_{gn} - \tau_g \simeq 0$.

## *2. Retrieval method for metamaterials*

With regard to the metamaterial with subwavelength elements, NRW Retrieval method is based on the assumption that we can describe a complex metamaterial within an a priori homogeneous model whose parameters may be retrieved from scattering measurement of a metamaterial sample. The retrieved parameters may be correctly used as an equivalent description of the metamaterial sample since NRW method may properly capture its propagation and scattering properties. However, this method has inherent limitations, usually associated with the initial assumption of simple homogenized models to describe complex metamaterial arrays, which cannot necessarily capture all the phenomena at the basis of their exotic properties. A more rigorous approach often requires considering high-order spatial dispersion effects that are not included in the conventional retrieval procedures.[77] A regular retrieval method for effective parameters of metamaterials proposed by Smith et al is introduced here.[78,79] The effective permittivity and permeability can be obtained via the refractive index ($n$) and impedance ($z$), which are related to S-parameters. S-parameters can be expressed by

$$S_{21} = \frac{1}{\cos(nkd) - \frac{i}{2}(z+\frac{1}{z})\sin(nkd)}, \tag{75}$$

$$S_{11} = \frac{i}{2}(\frac{1}{z} - z)\sin(nkd), \tag{76}$$

where $k$ is the wavenumber of the incident wave, and $d$ is the sample thickness. Eq. (75) and Eq. (76) can be inverted to find $n$ and $z$ in terms of the S-parameters as follow

$$n = \frac{1}{kd}\cos^{-1}\left[\frac{1}{2S_{21}}\left(1 - S_{11}^{2} + S_{21}^{2}\right)\right], \quad (77)$$

$$z = \pm\sqrt{\frac{(1+S_{11})^{2}-S_{21}^{2}}{(1-S_{11})^{2}-S_{21}^{2}}}. \quad (78)$$

The expressions of *n* and *z* complex functions with multiple branches lead to ambiguities in determining the final expressions for permittivity ($\varepsilon = n/z$) and permeability ($\mu = nz$). Additional knowledge about the material can resolve these ambiguities. For example, if the material is passive, the requirement of $\text{Im}(n) > 0$ and $\text{Re}(z) > 0$ can determine the correct value of *n* and *z*.

## C. Practical aspects as to how to implement the composites for specific EM applications

Having covered the fundamentals behind electromagnetic composites, we will now look at the design guidelines to implement them in real application platforms.

### 1. Electromagnetic shielding

There is a high demand for high-efficiency electromagnetic interference (EMI) shielding materials to protect the normal operation of electronic equipment, human health, and environment. Flexible materials with high conductivity that can provide EMI shielding with minimal thickness are highly looked-for, particularly if they can be easily processed into films that enable them to shield surfaces of any shape.[80] A highly efficient shield must both reduce undesirable emissions and protect the device from stray external signals. As the main function of shielding is to reflect radiation based on charge carriers and their interaction with EM fields, EMI shielding materials must be electrically conductive.[81] However, high conductivity is not the only requirement, shielding also requires absorption of the EM radiation through the material's electric and/or magnetic dipoles interacting with the incident wave. Coupling various types of conductive materials (such as carbon nanotube, carbon fiber, graphene, and MXene) with magnetic materials (Fe, Co, Ni, ferrite, and alloys) in polymer-matrix composites is a successful strategy for the dissipation of EM waves through promoting dielectric and magnetic losses.[82,83] A third mechanism that is less studied but also contributes to EMI shielding effectiveness corresponds to multiple internal reflections. Such internal reflections arise from interfaces or defects sites within the shielding material and thus a large surface area, porous, hollow and multiple shell structures are beneficial to promote them.[84]

## 2. Microwave absorption

Due to the high electrical conductivity of EMI shields, EM waves are mainly blocked by reflecting them into the outer space generating secondary EM pollution. Thus, EMI shielding materials with outstanding absorption must be also pursued. EM absorbers can attenuate more than 90% of electromagnetic waves if the reflection loss (RL) value (difference between the initial incident waves and the final reflected waves) is lower than -10 dB. The closer the values of $\varepsilon_r$ and $\mu_r$ are, the lesser the reflection is, and more reflection loss of EM waves can be achieved.[85] Therefore, excellent impedance matching is a key factor for high microwave absorption i.e. an effective complementarity between dielectric and magnetic losses should exist. Microstructural design should focus on improving both the impedance match and attenuation capacity. For example, meticulously regulating the pore size and distribution in aerogels and foams ensures that few EM waves are reflected at the surface owing to the lack of a continuous conductive surface and the presence of numerous insulated air pockets.[86] Another way to boost the microwave absorption is by assembling multilayers structures consisting of absorption and reflective layers. The absorption layer is an absorbing material with an excellent impedance match, and the reflective layer is composed of a highly conductive material with high shielding capacity. An electrically conductive gradient structure is therefore formed which provides an "absorption-reflection-reabsorption" path for EM waves of which absorption band and intensity can be tuned by adjusting the composition and thickness of the layers.[87-89] Ultra-broadband absorbers which are usually needed to e.g. cloak objects in the equivalent ultra-high frequency regime (from 300 MHz to 2 GHz) are usually thick, heavy or exhibit narrow absorption bandwidth.[90-92] Such issue can be resolved by a material–structure integrated design (MSID) methodology applied to dielectric metamaterials.[93] By combining e.g.

traditional honeycomb structures and periodic elements of metamaterials aided by numerical simulations, multiple resonances modes could be excited which is essential to the absorption performance.

*3. Antenna and radome applications*

Antenna components in telecommunication systems such as smart phones, tablets, and wireless internet devices are required to be more compact and capable of multiband operation. The antenna design considerations include miniaturization, gain, bandwidth, and efficiency.[94] To fulfill these requirements there are, e.g. microstrip antenna technologies miniaturized via high-permittivity dielectric substrates.[95,96] However, these antennas still exhibit narrow bandwidth and low gain. A novel solution for antenna design is the use of metamaterials which not only substantially reduces the antenna size but improves parameters such as bandwidth and gain. Basic antenna design can be in the form of a unit cell or multiple unit cells assembled into an array, so the first step is to design and study the parameters influencing resonance frequency $f_r$, permittivity, and permeability of the cell.[94] Depending on the structure and size of each unit cell, different $\varepsilon$, $\mu$, and $f_r$ are obtained. When metamaterials are used as part of the antenna structure, they should have high permeability values ($\mu \gg 1$) behaving as a magneto-dielectric substrate which reduces antenna size without using high permittivity substrates.[97,98] Metamaterials can be also used to overcome the low gain and improve bandwidth of small planar antennas by arranging unit cells around the radiation elements of the antenna[99] (Fig. 20a) or by placing the metamaterial on another dielectric layer called superstrate[100] (Fig. 20b). In the first case, the unit cells can be loaded on one side or both sides of the substrate, and their size must be investigated so that the resultant metamaterial matches the antenna $f_r$. Moreover, the unit cells must be easily integrated with radiated elements and used as insulators to reflect waves based on their negative $\mu$ characteristics. The gain is affected by both the number of unit cells and the $f_r$ of the designed antenna. For the second case, unit cells based on artificial magnetic conductors or artificial magnetic materials are loaded on one side or both sides of a superstrate. The power gain of the antenna will depend on the number of superstrate and unit cells and the distance between the radiation element and the superstrate.[101,102]. Finally, one should use a simulation tool in order to design and validate an antenna prior to fabrication using electromagnetic field solvers that can predict the behavior of radiating systems such as CST Microwave Studio. Not only optimizing antenna signal quality but also protecting antennas from harsh environmental conditions becomes essential to reliable and accurate antenna performance. The key to addressing this challenge lies in the radomes (acronym coined from radar dome) that surround and shield antennas. Ideally, the radome must allow transmission of EM waves and not damage the electrical performance of the enclosed antenna.[103] Moreover, structural and mechanical behavior from unexpected loads such as bird strike, air pressure, rain, and wind must be also considered.[104] The dielectric constant of the radome material should be lowered without compromising antenna protection. Multilayer or sandwich-structured radomes composed of a thick core layer of low permittivity and low-density material (such as honeycomb and foam structures) enwrapped by skin layers of fiber-reinforced composites are very attractive.[103-105] Designing foam materials with protective skins also address well the dielectric constant challenge.[106] With respect to mechanical considerations, these structures also have superiority compared to conventional laminates.[107,108] Both designs result in radomes with a known dielectric constant and signal reflection can be reduced or nearly eliminated by tuning the thickness profile and stacking sequences of the layers. Finally, to further optimize signal quality, absorbers can be used in the radome design[109].

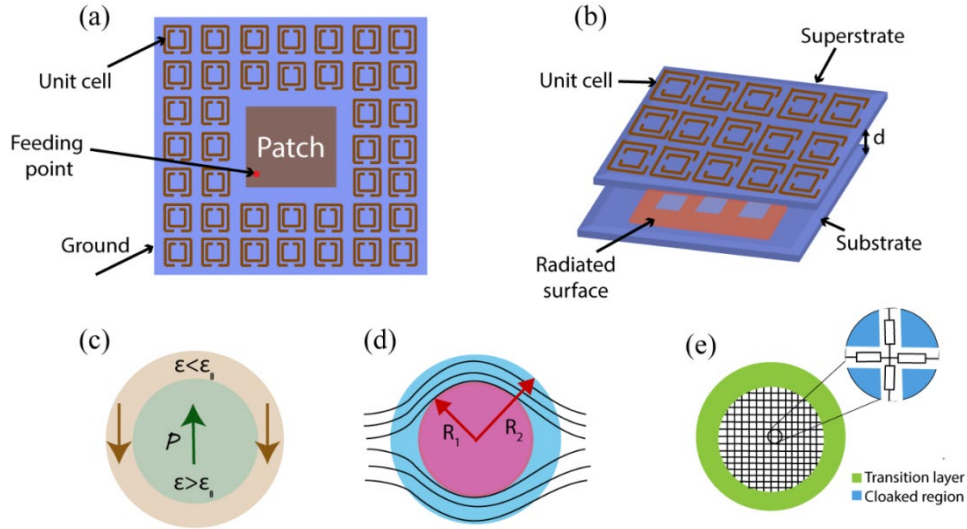

FIG. 20. Antenna design approaches: (a) Unit cells surrounding the radiated patch. (b) Placement of metamaterials on another dielectric layer called superstrate. Cloaking design approaches: (c) Scattering cancellation (d) Coordinate transformation; the electromagnetic field is guided inside the volume enclosed by the cloak. (e) Transmission line; the incident EM wave is coupled into a transmission-line network.

*4. Cloaking*

The design of electromagnetic cloaks is based on techniques such as scattering cancellation[110] (Fig. 20c), coordinate transformation[111] (Fig. 20d) and the use of metamaterials as dense meshes of transmission lines[112] (Fig. 20e). In the case of scattering cancellation, the main scattering object is covered by single or multiple layers of materials with relative permittivity $\varepsilon_r$ < 1 or plasmonic materials such as silver and gold.[113] The dielectric object (to be made "invisible") having permittivity larger than the surrounding medium can be covered with a dielectric shell having permittivity smaller than the surrounding medium. The shell thickness/diameter can be manipulated so that the scattering from the core and the shell cancel each other. For coordinate transformation, a spherical volume is designed where electromagnetic field does not exist but are rather guided around the volume. Cloaking the objects with this technique requires lossless anisotropic metamaterials with permittivity and permeability values smaller than these values in free space.[113] Finally, in the case of transmission lines, volumetric structures composed of 2D/3D transmission line networks are assembled.[112]

The electromagnetic field propagates inside the transmission lines leaving the volume between them efficiently cloaked.

## 5. *Structural health monitoring (SHM)*

Structural health monitoring (SHM) requires the development of integrated sensors arrays that are able to fulfil accuracy, robustness, low price, easy installment, simple operation, remote operation, and optimal sensor/structure bonding.[114]. Structural defects such as delamination, cracks, or inclusions can be sensed based on a local change in the electromagnetic properties.[115] In contrast to related techniques, electromagnetic non-destructive methods are non-contact, enable one-side scanning and are without safety concerns due to the low transmission power used.[116] Moreover, surface defects can be detected by reflection coefficient-based methods using waveguide probes and resonance frequency-based methods.[117,118] We and several researchers have demonstrated the potential of ferromagnetic microwire composites for SHM.[119-124] These composites constitute a system sensitive to stress thanks to their tunable properties derived from their giant magnetoimpedance (GMI) and giant stress impedance (GSI) effects.[125] The tensile stress in the matrix surrounding the microwires affects their response to an external magnetic field. The complex anisotropy distribution in these wires confers them with very fast domain propagation, which enables them to induce signals even from a small volume of microwires, allowing sensor miniaturization.[114] Negative Co-based microwires having a circumferential magnetized shell and longitudinally magnetized inner core are ideal for structural sensing applications. Their sensitivity can be further optimized by thermal, current, or stress annealing.[114] For instance, annealing temperatures between 200 to 375 °C and annealing times of 60 min are recommended to achieve a remarkable decrease in coercivity and improvement in the GMI effect. Application of stresses normally between 118 and 470 MPa during thermal treatment further improves magnetic softness and thus sensitivity. In addition, silane coupling agents can be also used to improve interfacial bonding with the matrix and improve the GSI sensitivity to meet application requirements for SHM.[120,121]

## VII. CONCLUSION

This tutorial deals with electromagnetic composites based on the treatment of two crucial electromagnetic constitutive parameters, permittivity and permeability. They are frequency-dependent and their frequency dispersion models including Debye, Lorentz, and Drude models elucidate the diverse responses of matter to electromagnetic wave arising from different fundamental origins. Aiming to resolve the relationship between these frequency dispersion properties and the microstructure of composites against a material science dimension, we first detailed the effective medium theories and computational simulation to describe and capture the statistically-average electromagnetic properties of composites satisfying certain predictions. Then we approach the intricate but fascinating structure-property relationship of composites from the effect of the key elements of composite microstructure, in particular the geometry and spatial dispersion of inclusions on the macroscopic properties significantly. It is demonstrated that it is possible to predict and design the electromagnetic composite properties via these microstructural factors. A special case predominantly dealing with periodical dispersion of inclusions is metamaterials with peculiar electromagnetic properties, such as negative refractive index, negative-permittivity (permeability), and zero refractive index, providing an innovative paradigm to tailor independently electric and magnetic responses to electromagnetic wave at the subwavelength dimension, which extend the scope of electromagnetic composites. Electromagnetic composites especially metamaterials have extensive application perspectives in microwave absorption or shielding, antennae, cloaking, and so on. Most importantly, the option of appropriate measurement methods and retrieval methods of constitutive parameters for different electromagnetic composites is the sine qua non for all the relevant research. With the knowledge acquired from this tutorial, it is anticipated that one could be ushered into the research and development of electromagnetic composites towards a diversity of application prospects. The future trend may lie in the AI aided programmable electromagnetic composites design, 3D and 4D printing smart fabrication and extended application of metamaterials with more advanced material-and-structure architecture in a spectrum of key industry sectors. Electromagnetic composites will be rising to these directions to inspire more young talents devoted into the field.

## ACKNOWLEDGEMENT

This work is supported by National Key Research and Development Program of China No. 2021YFB3501504, 2021YFE0100500 and ZJNSF No. LR20E010001 and Zhejiang Provincial Key Research and Development Program (2021C01004) and the Composites Platform Setup Fund by Ningbo Campus of Zhejiang University.

## DATA AVAILABILITY

The data that support the findings of this study are available from the corresponding author upon reasonable request.